\PassOptionsToPackage{hyphens}{url}
\documentclass[
11pt,
aps,
pra,
tightenlines,
longbibliography,
showpacs,
nofootinbib,
notitlepage,
superscriptaddress]{revtex4-2}

\usepackage[T1]{fontenc}
\usepackage{textcomp}
\usepackage{lmodern}     
\usepackage{microtype}
\usepackage{amsmath,amssymb,bm,mathtools}
\usepackage{graphicx}     
\usepackage{float}
\usepackage{booktabs,multirow}
\usepackage{listings}
\usepackage{xcolor}       
\usepackage{tikz}
\usetikzlibrary{positioning,shapes,arrows.meta,calc}

\usepackage{algorithm}
\usepackage[noend]{algpseudocode}

\newcommand*{\dif}{\mathop{}\!\mathrm{d}}

\usepackage{hyperref}

\begin{document}

\title{A Global Spacetime Optimization Approach to the Real-Space Time-Dependent Schrödinger Equation}

\author{Enze Hou}
\affiliation{Institute of Applied Physics and Computational Mathematics, Beijing 100094, China}
\affiliation{Graduate School of China Academy of Engineering Physics, Beijing 100088, China}

\author{Yuzhi Liu}
\affiliation{AI for Science Institute, Beijing 100080, China}

\author{Linxuan Zhang}
\affiliation{State Key Laboratory of Dark Matter Physics, Key Laboratory for Laser Plasmas (Ministry of Education) and School of Physics and Astronomy, Collaborative Innovation Center of IFSA (CICIFSA), Shanghai Jiao Tong University, Shanghai 200240, China}

\author{Difa Ye}
\affiliation{National Key Laboratory of Computational Physics, Institute of Applied Physics and Computational Mathematics, Fenghao East Road 2, Beijing 100094, China}

\author{Lei Wang}
\email{wanglei@iphy.ac.cn}
\affiliation{Institute of Physics, Chinese Academy of Sciences, Beijing 100190, China}

\author{Han Wang}
\email{wang\_han@iapcm.ac.cn}
\affiliation{National Key Laboratory of Computational Physics, Institute of Applied Physics and Computational Mathematics, Fenghao East Road 2, Beijing 100094, China}
\affiliation{%
HEDPS, CAPT, College of Engineering and School of Physics, Peking University, Beijing 100871, China
}

\begin{abstract}
The time-dependent Schrödinger equation (TDSE) in real space is fundamental to understanding the dynamics of many-electron quantum systems, with applications ranging from quantum chemistry to condensed matter physics and materials science.
However, solving the TDSE for complex fermionic systems remains a significant challenge, particularly due to the need to capture the time-evolving many-body correlations, while the antisymmetric nature of fermionic wavefunctions complicates the function space in which these solutions must be represented. 
We propose a general-purpose neural network framework for solving the real-space TDSE, Fermionic Antisymmetric Spatio-Temporal Network, which treats time as an explicit input alongside spatial coordinates, enabling a unified spatiotemporal representation of complex, antisymmetric wavefunctions for fermionic systems. 
This approach formulates the TDSE as a global optimization problem, avoiding step-by-step propagation and supporting highly parallelizable training. 
The method is demonstrated on five benchmark problems: a 1D harmonic oscillator, interacting fermions in a time-dependent harmonic trap, 3D hydrogen orbital dynamics, a laser-driven hydrogen atom, and a laser-driven H$_2$ molecule, achieving excellent agreement with reference solutions across all cases. 
These results demonstrate the method's accuracy and flexibility within the bound-state manifold across various dimensions and interaction regimes. 
While the current localized Ansatz inherently restricts the description of extensive ionization and continuum states, the method demonstrates the capability to stably simulate coherent multi-electron dynamics over extended time windows.
Our framework offers a highly expressive alternative to traditional basis-dependent or mean-field methods, opening new possibilities for ab initio simulations of time-dependent quantum systems, with applications in quantum dynamics, molecular control, and ultrafast spectroscopy.
\end{abstract}

\date{\today}

\maketitle

\section{Introduction}

The time-dependent Schrödinger equation (TDSE) serves as a cornerstone of quantum mechanics, governing the real-time evolution of quantum systems under time-dependent conditions. 
It plays a critical role across a wide range of physical contexts. 
These include ultrafast electronic dynamics in atomic and molecular systems, particularly high-order harmonic generation and multiphoton ionization, which underpin attosecond science~\cite{suarez2017high, tong1997theoretical}.
The contexts also encompass nonequilibrium phenomena in condensed matter physics, such as light-induced phase transitions and transient transport under strong laser fields~\cite{ibele2020excited, gao2019coulomb}.

Despite its fundamental importance, solving the TDSE for realistic quantum systems remains highly challenging. 
Analytic solutions exist only for a handful of simple models~\cite{griffiths2018introduction}, and numerical methods are indispensable for interacting many-body systems. 
However, numerically solving the TDSE in continuous real space remains extremely challenging due to the infinite-dimensional nature of the Hilbert space~\cite{lubich2008quantum}. 
The wavefunction's definition in such a functional space makes direct numerical computations highly demanding. 
Furthermore, accurately preserving essential quantum symmetries, most notably fermionic antisymmetry, during the time evolution further exacerbates the complexity~\cite{rubtsov2005continuous}.

To address the challenges of solving the TDSE in many-body systems, existing numerical approaches can be broadly understood in two directions.
First, direct real-space discretization methods (e.g., Crank-Nicolson~\cite{de1987product}, split-operator schemes~\cite{feit1982solution, hermann1988split}) evolve the wavefunction on a spatial grid by stepwise integration. 
While these schemes are systematically improvable and avoid physical-model approximations, they are intractable for realistic many-fermion systems due to the curse of dimensionality and the absence of explicit enforcement of essential physical constraints, such as fermionic antisymmetry~\cite{lubich2008quantum,bachmayr2016tensor}.
Second, quantum-chemistry methods introduce structured ansätze for the wavefunction, typically based on orbitals, Slater determinants, or truncated excitation expansions. 
Representative examples include real-time time-dependent density functional theory (RT-TDDFT)~\cite{runge1984density}, time-dependent Hartree-Fock (TDHF)~\cite{mclachlan1964variational}, multiconfiguration TDHF (MCTDHF)~\cite{meyer1990multi,beck2000multiconfiguration}, time-dependent configuration interaction (TDCI)~\cite{krause2005time}, and time-dependent coupled-cluster methods (TDCC)~\cite{sonk2011td, huber2011explicitly}. 
With these methods, the accuracy critically depends on the choice of the basis set and the quality of the initial guess.
Methods such as TDDFT and TDHF, which employ mean-field approximations to reduce the many-body problem to an effective single-particle framework, offer good computational efficiency and are well suited to weakly correlated or near-equilibrium regimes. 
However, their performance tends to degrade for strongly correlated systems or far-from-equilibrium dynamics~\cite{ullrich2011time, maitra2016perspective}.
TDCI and TDCC yield more accurate results, especially in capturing strong correlations, but their computational scaling remains steep. 
For instance, TDCCSD scales as $O(N^6)$, TDCCSDT as $O(N^8)$, and CISD as $O(N^6)$ per time step~\cite{sonk2011td, pathak2021time, hermann2023ab}, which limits their applicability to larger systems.

Despite their differences in wavefunction representation, both categories of methods ultimately rely on step-by-step time propagation using numerical integrators, such as Runge-Kutta integrators or Magnus/exponential propagators for grid-based solvers, and orbital/parameter propagation for quantum-chemistry approaches.
As a result, numerical errors accumulate over time and can severely degrade accuracy in long-time simulations.

In recent years, machine learning-based approaches have emerged as promising alternatives.
Neural-network quantum Monte Carlo (NN-QMC) methods based on the variational principle have demonstrated impressive performance in solving the time-independent Schrödinger equation (TISE), especially in modeling ground-state wavefunctions of fermionic systems.
By designing neural network ans\"{a}tze that incorporate physical constraints---such as permutation antisymmetry---researchers have achieved unprecedented accuracy in large-scale, strongly correlated electron systems. 
Neural network ans\"{a}tze have successfully modeled ground-state wavefunctions~\cite{han2019solving, pfau2020ab, hermann2020deep}, and more recently, have been extended to compute excited states~\cite{entwistle2023electronic, pfau2024accurate}, as reviewed in Ref.~\cite{hermann2023ab}.

Extending these successes to time-dependent problems, one major line of development employs the time-dependent variational principle (TDVP). 
Originally formulated for tensor networks~\cite{haegeman2011time}, this framework was adapted for neural quantum states in lattice models~\cite{carleo2017solving} and recently applied to real-space Hamiltonians~\cite{nys2024ab}.
While principled, TDVP-based simulators also propagate step by step, imposing stringent integrators and accumulating numerical errors due to the sequential nature of time evolution. 
Importantly, cumulative errors are fundamentally linked to the causality inherent in time evolution equations and therefore persist irrespective of numerical strategies.

In parallel, recent advancements have explored alternative strategies for solving partial differential equations (PDEs) in a global framework, such as by leveraging physics-informed neural networks (PINNs).
Following the original proposal~\cite{raissi2019physics}, the method has been enhanced with adaptive and parallel schemes~\cite{meng2020ppinn, wight2020solving}. 
Furthermore, sequential and pre-training strategies~\cite{mattey2022novel, prantikos2023physics, guo2023pre} have been introduced to provide effective initialization for subsequent time intervals, thereby mitigating optimization difficulties in complex evolution equations.
A key feature of PINN-based methods is their unsupervised nature, relying solely on the governing equations rather than external data labels.
These methods are primarily designed for solving general PDEs by embedding the PDE residuals and initial-boundary conditions into a loss function that is optimized over the entire spacetime domain, reducing the cumulative numerical errors associated with stepwise methods. 
However, while PINNs have demonstrated success in various contexts, the causal structure inherent in time-dependent equations remains a fundamental constraint, meaning that they can mitigate, but not entirely eliminate, cumulative errors~\cite{wang2022and, wang2024respecting}.

Building on the above PINN perspective, two spin-space works bring global-in-time residual objectives directly to TDSE.
The explicitly time-dependent neural quantum state (t-NQS)~\cite{van2025many} treats time as an input and optimizes over time windows; its loss is a TDSE residual minimized in spacetime, typically with autoregressive transformer architectures.
The time-dependent neural Galerkin (t-NQG) approach~\cite{sinibaldi2024time} uses a Galerkin-inspired ansatz, expanding the state as a linear combination of time-independent NQS with time-dependent coefficients, and enforces the TDSE globally by minimizing a projected residual.

Beyond these PINN-adjacent designs, two other spin-space formulations also pursue global-in-time training without stepwise propagation. 
Wang et al.~\cite{wang2021spacetime} introduce a self-normalized spacetime neural network architecture that couples an autoregressive Transformer for spatial modeling with a temporal network for time dependence.
As an early autoregressive treatment of time evolution in spin systems, it optimizes a global objective derived from an implicit-midpoint discretization and is demonstrated on imaginary-time Heisenberg models.
The smooth NQS (s-NQS)~\cite{wang2025continuous} makes the ansatz smooth and differentiable in time by expanding network parameters as linear combinations of temporal basis functions, and trains by maximizing a global fidelity objective that aligns the network's next-time state with the output of a short-time propagator acting on the current state.

All of the above global-in-time optimization works are formulated in discrete spin-space, so explicit continuous real-space antisymmetry is not required.
To the best of our knowledge, no work has demonstrated a global spacetime optimization of the real-space TDSE for interacting fermions, where an explicitly antisymmetric wavefunction in continuous coordinates is essential, and existing PINN-style frameworks have not yet addressed this setting.

This work explores an alternative to conventional stepwise propagation. 
We introduce a general antisymmetric neural-network wavefunction defined in continuous real space with time as an explicit input coupled to spatial coordinates. 
This architecture enables a unified framework for representing time-dependent solutions of fermionic systems with arbitrary interaction forms, without redesigning the network structure for each new problem.
By formulating the TDSE as a global spacetime optimization problem with a residual loss, our approach avoids explicit ODE integration and enables highly parallelized training. 
Although the intrinsic causal structure of time evolution implies that error propagation cannot be entirely eliminated, our results demonstrate that accurate long-time simulations are feasible within this framework.
Rather than replacing stepwise solvers, this method provides a complementary route to real-space TDSE solutions that is scalable, expressive, and generalizable, moving beyond the step-by-step paradigm.


\section{Methodology}

In this work, we propose a neural-network-based method to solve the TDSE that describes the evolution of a quantum many-electron system defined on a continuous, open spatial domain. 
The governing equation is formulated as follows: 
\begin{equation}\label{eq:tdse}
    i \frac{\partial \Psi(\bm{X}, t)}{\partial t} = \hat{H}(t)\Psi(\bm{X}, t),\quad \bm{X}\in\Omega, t\in[0,T],
\end{equation}
It is subject to the initial condition:
\begin{equation}\label{eq:initial_condition}
\Psi(\bm{X}, 0) = \Psi_0(\bm{X}),\quad \bm{X}\in\Omega.
\end{equation}

Here, the full configuration $\bm{X} = \{\bm{x}_1, \bm{x}_2, ..., \bm{x}_N\}$ lies in the domain $\Omega = (\mathbb{R}^{d} \times \{\uparrow, \downarrow\})^{\times N}$, where $d$ is the spatial dimension and $N$ denotes the number of electrons. 
Each electron coordinate $\bm{x}_i = \{\bm{r}_i, \sigma_i\}$ consists of a spatial coordinate $\bm{r}_i \in \mathbb{R}^d$ and a spin variable $\sigma_i \in \{\uparrow, \downarrow\}$.
Thus, the wavefunction $\Psi(\bm{X}, t)$ is defined over the Cartesian product of the continuous spatial domain $\mathbb{R}^{dN}$, the discrete spin domain, and the temporal domain $[0,T]$.
The Hamiltonian governing the system dynamics takes the following form:
\begin{equation}
    \hat{H}(t) = -\frac{1}{2} \sum_{i=1}^N \nabla^2_{\bm{r}_i} + V(\bm{X}, t), 
\end{equation}
where $\nabla^2_{\bm{r}_i}$ denotes the Laplacian operator acting on the position of the $i$-th electron, and $V(\bm{X}, t)$ represents a time-dependent potential, which may include external fields and electron-electron, electron-nucleus interactions.
Throughout this paper, atomic units are used for simplicity.

Since the Hamiltonian considered here is spin-independent, each electron's spin state remains unchanged throughout the time evolution. 
Thus, without loss of generality, we adopt a fixed spin-assignment strategy to simplify the computational approach.
Specifically, we assume that the first $N_{\uparrow}$ electrons are spin-up, and the remaining $N_{\downarrow} = N - N_{\uparrow}$ electrons are spin-down.
This assumption allows us to omit explicit spin variables and simplify the wavefunction to $\Psi(\bm{r}, t)$, where $\bm{r} = \{\bm{r}_1, \bm{r}_2, ..., \bm{r}_N\}$ denotes only the spatial coordinates.

According to Fermi-Dirac statistics, fermions obey the Pauli exclusion principle, which states that no two fermions can simultaneously occupy the same quantum state.
As fermions, electrons have a wavefunction that is antisymmetric under the exchange of any two electrons with the same spin:
\begin{equation}
    \Psi(..., \bm{r}_i, ..., \bm{r}_j, ..., t) = -\Psi(..., \bm{r}_j, ..., \bm{r}_i, ..., t), \quad
    \forall\ 0 < i < j \leq N_{\uparrow} \ \text{or}\ N_{\uparrow} < i < j \leq N.
\end{equation}
To approximate the wavefunction \(\Psi(\bm{r}, t)\), we use a neural network ansatz that explicitly takes the spatial coordinates and time as inputs.
This ansatz enforces essential physical constraints, including fermionic antisymmetry and translational invariance.
The neural network is trained by minimizing the TDSE residuals~\eqref{eq:tdse} and the initial condition.
The following sections present the neural-wavefunction architecture, the loss function, and the optimization strategy.

\subsection{Neural Ansatz: FASTNet}

We represent the time-dependent wavefunction $\Psi(\bm{r}, t)$ using a neural network architecture inspired by FermiNet~\cite{pfau2020ab}, introducing key innovations to accommodate real-time quantum dynamics.
Specifically, our design incorporates explicit time inputs throughout the network to enable joint spatiotemporal modeling, and it extends the architecture with time-evolving envelopes and a trainable complex phase factor. 
We refer to the resulting architecture as FASTNet (Fermionic Antisymmetric Spatiotemporal Network), highlighting its capability to efficiently and accurately represent complex-valued, antisymmetric, time-dependent fermionic wavefunctions in continuous space.

\begin{figure}
    \centering
    \includegraphics[width=1.0\textwidth]{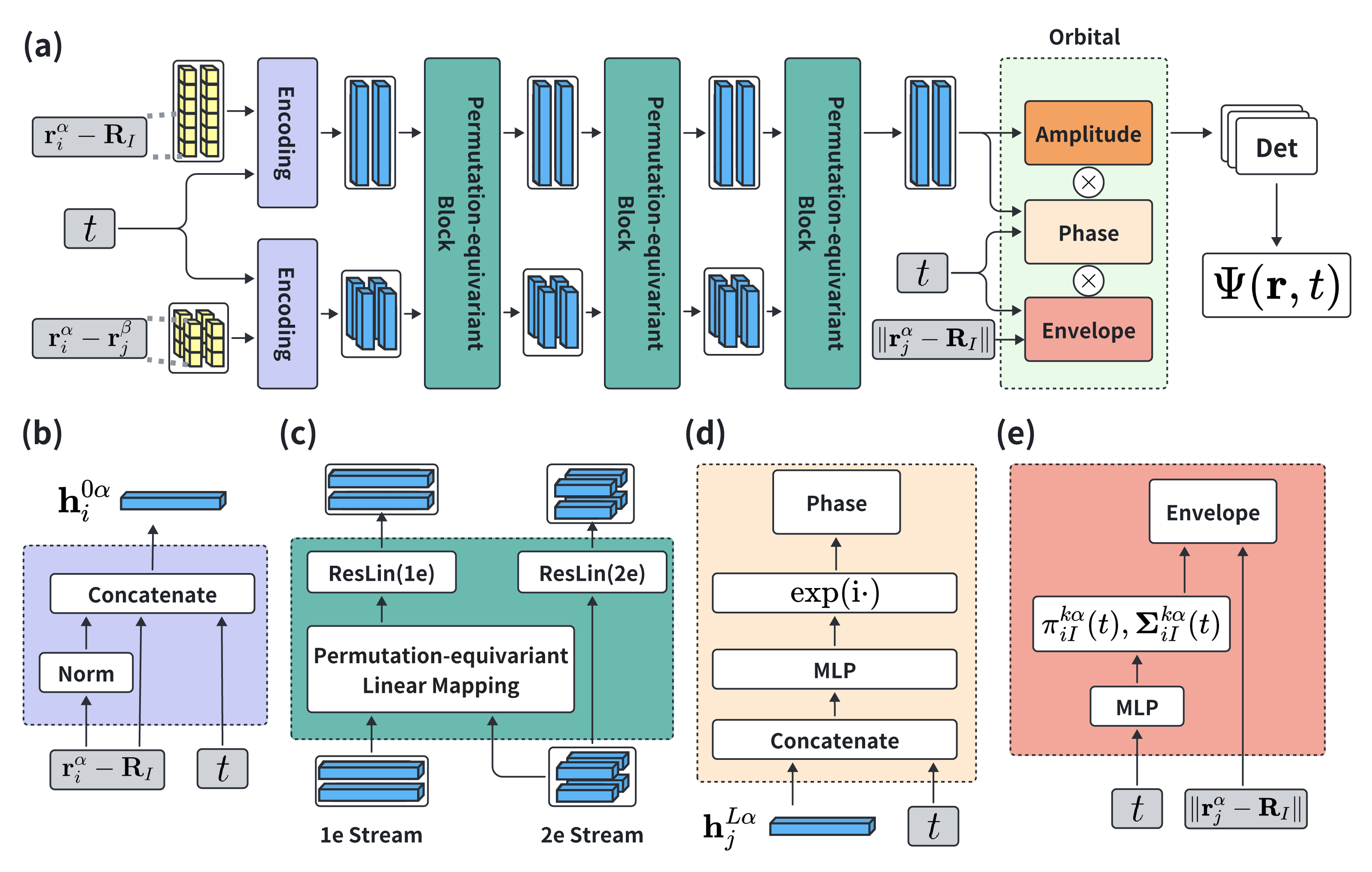}
    \caption{
        Architecture of FASTNet for time-dependent electronic wavefunctions.
        (a) Overview of the network structure. 
        (b-e) Detailed views of the embedding, permutation-equivariant block, phase, and envelope modules, respectively.
    }
    \label{fig:FASTNet_arch_corrected}
\end{figure}

\paragraph{Spatiotemporal Input Encoding.}
To effectively encode spatial and temporal dependencies, we construct one-electron and two-electron features by concatenating relative position vectors, pairwise distances, and time:
\begin{equation}
\begin{aligned}
    & \text{One-electron stream:} \quad \bm{h}_i^{\ell\alpha} = \left[ \bm{r}_i^\alpha - \bm{R}_I,\ \|\bm{r}_i^\alpha - \bm{R}_I\|,\ t \right]_{\forall I}, \\
    & \text{Two-electron stream:} \quad \bm{h}_{ij}^{\ell\alpha\beta} = \left[ \bm{r}_i^\alpha - \bm{r}_j^\beta,\ \|\bm{r}_i^\alpha - \bm{r}_j^\beta\|,\ t \right]_{\forall j,\beta},
\end{aligned}
\end{equation}
where $\alpha, \beta$ denote spin states, and $I$ indexes the nuclei. 
These time-aware encodings are propagated through multiple layers to enable dynamic backflow transformations.

\paragraph{Permutation-Equivariant Block.}
Building upon the FermiNet architecture, we implement a permutation-equivariant block that updates the one- and two-electron feature streams at each intermediate layer, facilitating structured information exchange and feature transformations among electrons.
At each layer $\ell$, the one-electron features $\bm{h}_i^{\ell\alpha}$ are updated using spin-resolved mean activations aggregated from the one- and two-electron streams.
Specifically, for each electron $i$ with spin $\alpha$, we compute:
\begin{equation}
\text{Permutation-equivariant linear mapping:} \quad \bm{f}_i^{\ell\alpha} = \left( \bm{h}_i^{\ell\alpha},\ \bm{g}^{\ell\uparrow},\ \bm{g}^{\ell\downarrow},\ \bm{g}_i^{\ell\alpha\uparrow},\ \bm{g}_i^{\ell\alpha\downarrow} \right),
\end{equation}
where the pooled mean activations are defined as $\bm{g}^{\ell\beta} = \frac{1}{n^\beta} \sum_j \bm{h}_j^{\ell\beta}$ and $\bm{g}_i^{\ell\alpha\beta} = \frac{1}{n^\beta} \sum_j \bm{h}_{ij}^{\ell\alpha\beta}$. 
This concatenated vector $\bm{f}_i^{\ell\alpha}$ is then passed through a residual linear layer with variance normalization (ResLin)—i.e., a single affine transformation with a skip connection scaled by $1/\sqrt{2}$—to produce the updated one-electron features:
\begin{equation}
    \text{ResLin (1e):}\quad 
    \bm{h}_i^{\ell+1,\alpha} \;=\; \frac{1}{\sqrt{2}}\left( \tanh\!\left( \bm{V}^\ell \bm{f}_i^{\ell\alpha} + \bm{b}^\ell \right) \;+\; \bm{h}_i^{\ell\alpha} \right),
\end{equation}
Analogously, the two-electron stream is updated element-wise by:
\begin{equation}
\text{ResLin (2e):} \quad \bm{h}_{ij}^{\ell+1, \alpha\beta} \;=\; \frac{1}{\sqrt{2}}\left( \tanh\!\left( \bm{W}^\ell \bm{h}_{ij}^{\ell\alpha\beta} + \bm{c}^\ell \right) \;+\; \bm{h}_{ij}^{\ell\alpha\beta} \right).
\end{equation}
This $1/\sqrt{2}$ residual scaling helps stabilize the feature variance across layers.

Together, these operations constitute a permutation-equivariant block.
By stacking multiple such blocks, the network progressively learns deep representations that are equivariant under permutations of electrons within each spin channel, enabling the architecture to encode symmetry-aware interactions across particles.

\paragraph{Time-Dependent Envelopes.}
To model the evolving asymptotic behavior under external potentials, we employ time-conditioned envelope functions. 
For each electron-nucleus pair $(i, I)$, the envelope parameters $\pi_{iI}^{k\alpha}(t)$ and $\bm{\Sigma}_{iI}^{k\alpha}(t) \in \mathbb{R}^{3\times 3}$ are treated as learnable, time-dependent functions.
We adopt the following unified form: 
\begin{equation}
    \mathrm{Env}^{k\alpha}_{i}(\bm{r}_j^\alpha, t) 
    = \sum_I \pi_{iI}^{k\alpha}(t) 
    \exp\!\left(-\|\bm{\Sigma}_{iI}^{k\alpha}(t)\left(\bm{r}_j^\alpha - \bm{R}_I\right)\|^{p}\right),
\end{equation}
where the exponent $p > 0$ is a structural hyperparameter determined by the physical potential (e.g., $p=1$ for Coulombic decay). 
Crucially, while $p$ fixes the asymptotic form, the learnable parameters $\pi_{iI}^{k\alpha}$ and $\bm{\Sigma}_{iI}^{k\alpha}$ allow the envelope to evolve dynamically. 
This flexibility enables the ansatz to accurately represent various bound configurations, including spatially diffuse excited states. 
We note, however, that this ansatz enforces spatial localization, making it well-suited for bound-state dynamics and recollision processes, but not for unbound continuum states with significant ionization flux.

\paragraph{Phase Factor.}
The complex phase evolution of the wavefunction is modeled by a trainable phase factor applied to each electron, thereby introducing a time-dependent modulation. 
Specifically, the $i$-th orbital wavefunction is multiplied by a factor evaluated at the $j$-th electron with spin $\alpha$ at time $t$: 
\begin{equation}
J_i^{k\alpha}(\bm{r}_j^\alpha, t) = \exp\left[\mathrm{i}\, S^{i k}_\theta(\bm{h}_j^{L\alpha}, t)\right],
\end{equation}
where $S^{i k}_\theta$ is a two-layer MLP that takes the final-layer one-electron stream $\bm{h}_j^{L\alpha}$ concatenated with the explicit time variable $t$ as input.
The output $S^{i k}_\theta(\cdot)$ is a real scalar representing the phase angle $\theta$, and the exponential mapping $J_i^{k\alpha}(\cdot)$ ensures that the resulting factor lies on the unit circle in the complex plane. 
This architecture enables the model to capture smooth, time-dependent variations in the quantum phase, which are essential for accurately representing nonstationary dynamics in quantum systems.

\paragraph{Base Amplitude.}
The base-amplitude component is an intermediate network output that serves as the foundational spatial feature of the wavefunction prior to envelope modulation and phase modulation. 
Specifically, we obtain the base amplitude by applying a linear projection to the final one-electron-stream outputs, capturing essential spatial features required to model electron correlation:
\begin{equation}
\text{Base amplitude} = \bm{w}_i^{k\alpha} \cdot \bm{h}_j^{L\alpha} + g_i^{k\alpha}.
\end{equation}
This intermediate representation is subsequently modulated by time-dependent envelopes and complex phase factors to construct the complete orbital wavefunction.

\paragraph{Time-Evolving Orbitals.}
The final orbital wavefunction is constructed by combining the base amplitude, envelope, and phase components:
\begin{equation}
\phi_i^{k\alpha}(\bm{r}_j^\alpha, t) = \underbrace{\left(\bm{w}_i^{k\alpha} \cdot \bm{h}_j^{L\alpha} + g_i^{k\alpha}\right)}_{\text{Base Amplitude}} \cdot
\underbrace{\mathrm{Env}^{k\alpha}_{i}(\bm{r}_j^\alpha,t)}_{\text{Envelope}} \cdot
\underbrace{J_i^{k\alpha}(\bm{r}_j^\alpha,t)}_{\text{Phase}}.
\end{equation}
Notably, all components except for the phase are real-valued, and antisymmetry is enforced via Slater-determinant construction within each spin channel. 
Final output of FASTNet is expressed as a weighted average of full Slater determinants of orbitals:
\begin{equation}
  \Psi_\eta(\bm{r}, t)=\sum_k \omega_k \operatorname{det}\left[\phi_i^{k}(\bm{r}_j, t)\right],
\end{equation}
where $\eta$ denotes the complete set of trainable parameters in the wavefunction ansatz.
The complete network-evaluation algorithm and computational cost analysis can be found in \hyperref[alg:ferminet]{Appendix A}.
The architecture of FASTNet is shown in Figure~\ref{fig:FASTNet_arch_corrected}.

\subsection{Loss Function and Training Procedure}

We formulate the TDSE as a spacetime residual minimization problem. 
Given an initial wavefunction $\Psi_0(\bm{r})$ and a time-dependent Hamiltonian $\hat{H}(t)$, the system evolves according to the time-dependent Schrödinger equation and the initial condition, as specified in Eqs.~\eqref{eq:tdse}-\eqref{eq:initial_condition}.
The initial condition can be provided in analytical form or obtained through VMC methods (in this work we use FermiNet) to represent a specific ground state or excited state.

Notably, we do not explicitly enforce wavefunction normalization in our formulation. 
Instead, we require the wavefunction to satisfy the TDSE; for a Hermitian Hamiltonian, this implies that the norm is conserved over time. 
When computing observables through MCMC sampling, any global normalization factor cancels out, making this approach both efficient and mathematically sound.

Defining the local energy as $E_L(\bm{r}, t) = \dfrac{\hat{H}\Psi_\eta}{\Psi_\eta}(\bm{r}, t)$, we construct a strong-form loss function over the spacetime domain $\mathbb{R}^{d \times N} \times [0,T]$, consisting of two components: the residual loss $\mathcal{L}_R$ enforcing the TDSE, and the initial loss $\mathcal{L}_I$, enforcing the initial condition. 
The total loss is
\begin{equation}\label{eq:total_loss}
    \mathcal{L}(\eta;\mathcal{T}) = \lambda_R \mathcal{L}_R(\eta;\mathcal{T}_R) + \lambda_I \mathcal{L}_I(\eta;\mathcal{T}_I),
\end{equation}
with
\begin{equation}\label{eq:residual_loss}
\begin{aligned}
    \mathcal{L}_R(\eta;\mathcal{T}_R) &= \mathbb{E}_{p_{\eta}(\bm{r}, t)} \left[ \left|i \frac{\partial \log \Psi_\eta}{\partial t}(\bm{r}, t) - E_L(\bm{r}, t) \right|^2 \right], \\
    \mathcal{L}_I(\eta;\mathcal{T}_I) &= \mathbb{E}_{p_0(\bm{r})} \left[ \left| \Psi_\eta(\bm{r}, 0) - \Psi_0(\bm{r}) \right|^2 \right].
\end{aligned}
\end{equation}
Here, $\lambda_R$ and $\lambda_I$ control the relative weight of minimizing the residual of the TDSE and satisfying the initial condition.

The data sets $\mathcal{T}_R$ and $\mathcal{T}_I$ are
\begin{equation}
    \mathcal{T}_R = \left\{ (\bm{r}^j, t^j) \right\}_{j=1}^{N_R}, 
    \quad 
    \mathcal{T}_I = \left\{ \bm{r}^j \right\}_{j=1}^{N_I},
\end{equation}
with $(\bm{r}^j, t^j) \in \mathbb{R}^{d\times N} \times [0,T]$ sampled from a spatiotemporal probability measure $p_{\eta}(\bm{r}, t)$, and $\bm{r}^j$ drawn from the initial probability measure $p_0(\bm{r}) = |\Psi_0(\bm{r})|^2 / \langle \Psi_0|\Psi_0\rangle$.
Note that these are configuration points used to estimate the residual loss and initial loss, not labeled data from external simulations.

A key computational insight is that the expectation in $\mathcal{L}_R$ is evaluated via the conditional decomposition:
\begin{equation}
\mathbb{E}_{p_{\eta}(\bm{r}, t)}[\,\cdot\,] 
= \mathbb{E}_{p(t)}\!\left[\, \mathbb{E}_{p_{\eta}(\bm{r}|t)}[\,\cdot\,] \right],
\end{equation}
where $p(t)$ is uniform over $[0,T]$ and $p_{\eta}(\bm{r}|t)$ represents the spatial sampling distribution at fixed time $t$.
This decomposition strategy is computationally advantageous: direct sampling from the joint distribution $p_{\eta}(\bm{r}, t) \propto |\Psi_\eta(\bm{r},t)|^2$ would require evaluating the time-dependent normalization factor $\langle \Psi_\eta|\Psi_\eta\rangle(t)$ at each time point, involving costly high-dimensional integrals. 
Instead, we discretize the time domain into $N_t$ uniform steps, and at each fixed $t$ perform Monte Carlo sampling from
\begin{equation}
    p_{\eta}(\bm{r}|t)=|\Psi_{\eta}(\bm{r},t)|^2/\langle \Psi_{\eta}|\Psi_{\eta}\rangle(t).
\end{equation}
In practice, Metropolis-Hastings updates rely only on the ratio of probabilities between successive proposals, so the normalization factor in the denominator cancels out.
This property allows us to draw residual points without explicitly computing $\langle \Psi_{\eta}|\Psi_{\eta}\rangle(t)$.
The same VMC framework is applied for both initial sampling and fixed-time residual sampling.


Since the residual loss is an expectation over a parameterized distribution, the gradient is nontrivial. 
An unbiased estimate of its gradient yields:
\begin{equation}
\begin{aligned}
\nabla_\eta \mathcal{L}_{R} &= \mathbb{E}_{p_{\eta}(\bm{r}, t)} \left[ (\nabla_\eta \log \Psi_\eta^* + \nabla_\eta \log \Psi_\eta) \cdot \left| i\frac{\partial \log \Psi_\eta}{\partial t} - E_L \right|^2 \right] \\
&\quad - \mathbb{E}_{p(t)} \left[ \mathbb{E}_{p_{\eta}(\bm{r}\vert t)} \left[ \nabla_\eta \log \Psi_\eta^* + \nabla_\eta \log \Psi_\eta \right] \cdot \mathbb{E}_{p_{\eta}(\bm{r}\vert t)} \left[ \left| i\frac{\partial \log\Psi_\eta}{\partial t} - E_L \right|^2 \right] \right] \\
&\quad + \mathbb{E}_{p_{\eta}(\bm{r},t)} \left[ \nabla_\eta \left| i\frac{\partial \log\Psi_\eta}{\partial t} - E_L \right|^2 \right].
\end{aligned}
\end{equation}
Derivations are provided in the Appendix Sec.~\ref{sec:loss-gradient}.

Training is conducted in two stages. 
In the first stage, we use the Adam optimizer with a linear warmup followed by exponential decay of the learning rate. 
At each iteration, MCMC sampling is performed to update the residual and initial points, enhancing robustness against sampling noise. 
In the second stage, we switch to the L-BFGS optimizer to fine-tune the network. 
Since the sampling distribution evolves as optimization proceeds and L-BFGS requires a fixed dataset during optimization,  we periodically resample a fresh batch of points after a fixed number of L-BFGS steps.
After each resampling, the optimizer is restarted to ensure stable convergence with the updated data distribution.
Throughout both stages, we apply gradient clipping to mitigate exploding gradients and improve convergence stability. 
All training is conducted in double precision and supports full multi-GPU parallelism across time slices, and we use Forward Laplacian method~\cite{li2024computational} to compute second-order derivatives efficiently.

\subsection{Pretraining Strategy for Long-Time Evolution}

\paragraph{Motivation.}
Accurately simulating long-time quantum dynamics with neural networks is challenging due to the highly non-convex loss landscape associated with large parameter spaces, and the vast solution space induced by complex time-dependent Hamiltonians.
Our approach builds on the pretraining paradigm: instead of training over the full temporal domain from the outset, we solve a sequence of shorter, overlapping time subproblems, progressively propagating information forward in time.

This strategy alleviates the difficulty of optimization, since training over limited subintervals is substantially more tractable\cite{guo2023pre, mattey2022novel, wight2020solving}, particularly in regimes with rapid temporal variations. Moreover, when the solution to the underlying PDE is smooth in time, the previous interval's network naturally serves as an effective initialization for the next due to extrapolation property\cite{guo2023pre, karniadakis2021physics}, allowing the method to exploit the inherent extrapolation capability of the learned representation.

In the proposed pretraining scheme, the full time domain is divided into overlapping subintervals. 
The first interval is trained from scratch using only the initial condition; subsequent intervals are initialized from the previous interval's parameters and trained with both the PDE residual and continuity penalties at the overlap. 
This procedure can be applied with uniform partitions, or adaptively when prior knowledge of the Hamiltonian suggests where finer temporal resolution is required.

\paragraph{Time-Partition.}
We divide the time domain $[0, T]$ into $M$ subintervals.  
If prior information on the variation intensity of $\hat{H}(t)$ is available, we adaptively assign shorter intervals to regions with rapid variation and longer intervals to smoother regions; otherwise, we use a uniform partition.

Each subinterval overlaps with the next by $5\%$ of its length.  
This serves two purposes: (i) during training of the preceding interval, sampling points on both sides of the temporal boundary allows accurate evaluation of derivatives—particularly time derivatives—directly at the boundary, and (ii) it stabilizes optimization for the next interval by starting from a region already well learned in the previous one.  
The resulting overlaps are:
\begin{equation}
[t_0, t_1 + 0.05\Delta t_1],\ [t_1, t_2 + 0.05\Delta t_2],\ \dots,\ [t_{M-1}, t_M + 0.05\Delta t_M],
\end{equation}
where $t_0=0$, $t_M=T$, and $\Delta t_i = t_i - t_{i-1}$.

\paragraph{Training Procedure Per Interval.}
For the first interval $[t_0, t_1+0.05\Delta t_1]$, we minimize the loss in Eq.~\eqref{eq:total_loss} using residual and initial terms:
\begin{equation}
\eta^{(1)} = \arg\min_{\eta} \mathcal{L}^{(1)}(\eta;\mathcal{T}^{(1)}).
\end{equation}
For the subsequent intervals $[t_{i-1}, t_i+0.05\Delta t_i]$, $i\ge2$, we initialize
\begin{equation}
\eta^{(i)}_{\mathrm{init}} \leftarrow \eta^{(i-1)},
\end{equation}
and train using:
\begin{equation}
\mathcal{L}^{(i)} = \lambda_R \mathcal{L}_R + \lambda_{PV}\mathcal{L}_{PV} + \lambda_{PT}\mathcal{L}_{PT} + \lambda_{PS}\mathcal{L}_{PS},
\end{equation}
where $\mathcal{L}_R$ is the residual loss over the new interval, and the penalty terms $\mathcal{L}_{PV}$, $\mathcal{L}_{PT}$ and $\mathcal{L}_{PS}$ enforce $\ell_2$ continuity in $\Psi$, $\partial_t\Psi$ and $\nabla_{\bm{r}}\Psi$, respectively, at $t=t_{i-1}$, with boundary samples drawn from fixed probability distribution $p(\bm{r}|t_{i-1}) \propto |\Psi_{\eta^{(i-1)}}(\bm{r},t_{i-1})|^2$.
The penalty terms are explicitly defined as:
\begin{equation}
\begin{aligned}
\mathcal{L}_{PV} &= \mathbb{E}_{\bm{r}\sim p(\bm{r}|t_{i-1})} 
    \big[|\Psi_{\eta^{(i)}}(\bm{r},t_{i-1}) - \Psi_{\eta^{(i-1)}}(\bm{r},t_{i-1})|^2\big], \\
\mathcal{L}_{PT} &= \mathbb{E}_{\bm{r}\sim p(\bm{r}|t_{i-1})} 
    \big[\|\partial_t\Psi_{\eta^{(i)}}(\bm{r},t_{i-1}) - \partial_t\Psi_{\eta^{(i-1)}}(\bm{r},t_{i-1})\|_2^2\big], \\
\mathcal{L}_{PS} &= \mathbb{E}_{\bm{r}\sim p(\bm{r}|t_{i-1})} 
    \big[\|\nabla_{\bm{r}}\Psi_{\eta^{(i)}}(\bm{r},t_{i-1}) - \nabla_{\bm{r}}\Psi_{\eta^{(i-1)}}(\bm{r},t_{i-1})\|_2^2\big].
\end{aligned}
\end{equation}
Note that minimizing these Euclidean norms for the complex-valued wavefunction inherently constrains both the magnitude and the phase, thereby preventing unphysical phase discontinuities at the temporal boundaries.

\begin{figure}
    \centering
    \includegraphics[width=\textwidth]{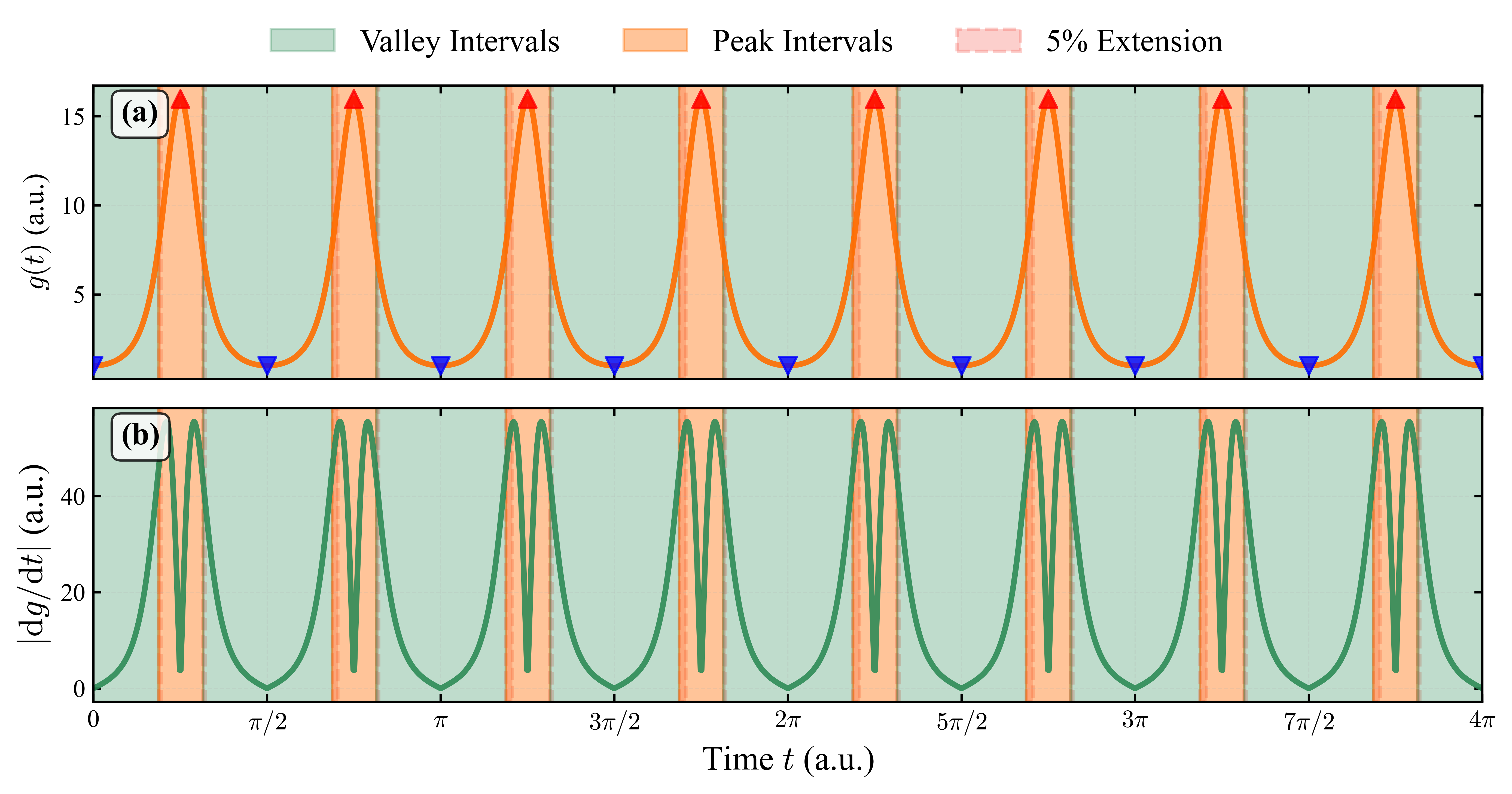}
    \caption{Illustration of the pre-training strategy. 
      The time domain is divided into overlapping intervals, with each interval trained using the previous interval's parameters as initialization. 
      The overlap ensures smooth transition between intervals, allowing for efficient and accurate training of long-time dynamics.
    }
    \label{fig:time_partition}
\end{figure}

\paragraph{Illustrative Example.}
We demonstrate the method on the setting of \hyperref[sec:1d-interaction]{Section~3.2}, where after the quench the only time-dependent parameter is the interaction strength $g(t)$, which is periodic with period $\pi/2$.
Within each period, the “peak” regions (near the maxima of $g(t)$) exhibit rapid variations and are therefore covered with shorter subintervals ($\Delta t = 0.4$), whereas the smoother “valley” regions (near the minima) admit longer subintervals ($\Delta t \approx 1.17$).
In both regimes, each subinterval is extended by 5\% to overlap with its successor, ensuring continuity across boundaries. 
The resulting partition of $[0,12]$ is illustrated in Figure~\ref{fig:time_partition}.

\paragraph{Discussion.}
Our method draws inspiration from the PT-PINN method\cite{guo2023pre}, and can be positioned within the broader family of sequential or transfer-learning strategies for PINNs. 
Compared with PT-PINN, which gradually enlarges the training interval using a single network, we explicitly partition the time domain into overlapping subintervals, each with its own set of parameters initialized from the previous segment. 
This avoids forcing one network to represent the entire temporal evolution, which is particularly beneficial for many-body quantum dynamics where the relevant time horizon is much longer and more complex.

Our strategy is also closely related to time-marching approaches. 
For example, Wight and Zhao~\cite{wight2020solving} expand the training domain adaptively based on the residual loss, while Mattey and Ghosh~\cite{mattey2022novel} partition the domain into subintervals and enforce continuity and consistency with previously learned segments to avoid loss of earlier information. 
Both methods employ a single neural network to cover the entire temporal range. 
Methods based on loss reweighting, such as curriculum regularization~\cite{krishnapriyan2021characterizing} and the causality-enforcing reformulation~\cite{wang2024respecting}, follow a different philosophy. 
They retain a single global network and encode temporal causality implicitly through the optimization objective. 
By contrast, our approach enforces causality and continuity structurally: parameter transfer across subintervals provides a natural initialization, while overlaps supply direct supervision at interfaces. 
This explicit decomposition is particularly advantageous for long-time quantum dynamics, where a single network often lacks sufficient capacity to capture the entire evolution.

We note that, as with other pretraining strategies, numerical errors in our method still accumulate over time, and we do not fundamentally eliminate this issue. 
The benefit lies instead in making each subproblem more tractable to train, thereby improving stability and accuracy on intermediate intervals and, in practice, extending the feasible horizon of neural-network-based quantum dynamics simulations.



\section{Results}

To assess the effectiveness and generality of FASTNet, we conduct experiments on five representative quantum systems that differ in physical characteristics and dimensionality.
These include:
(1) 1D single harmonic oscillator,
(2) interacting fermions in 1D harmonic trap,
(3) the real-time evolution of hydrogen atomic orbitals in 3D, 
(4) 3D hydrogen atom in a strong elliptically polarized laser field, and
(5) H$_2$ molecule driven by weak and strong linearly polarized laser fields.
These test cases span analytically solvable models, strongly correlated many-body systems, and high-dimensional real-space problems, allowing us to systematically assess the accuracy, expressivity, and scalability of our method across diverse dynamical regimes.

\subsection{1D Single Harmonic Oscillator}

We first validate FASTNet on a prototypical solvable system: a single particle in a 1D harmonic potential, governed by the time-dependent Schrödinger equation:
\begin{equation}
    \hat{H} = -\frac{1}{2} \frac{\partial^2}{\partial r^2} + \frac{1}{2} m \omega^2 r^2,
\end{equation}
with $m = 1$ and $\omega = 1$, where $r$ denotes the position coordinate.

The analytical solution of this system is well established.
For an initial state expressed as a linear combination of eigenstates, 
\begin{equation}
\Psi(r, 0) = \sum_{n \in \mathcal{N}} c_n \psi_n(r),
\end{equation}
where $\psi_n(r) = N_n H_n(\sqrt{m\omega}r)\, e^{-\frac{m\omega r^2}{2}}$ is the $n$-th eigenstate with energy $E_n = \omega(n + \frac{1}{2})$, the time evolution is given by
\begin{equation}
\Psi(r, t) = \sum_{n \in \mathcal{N}} c_n \psi_n(r) e^{-i E_n t}.
\end{equation}

\begin{figure}
    \centering
    \includegraphics[width=\textwidth]{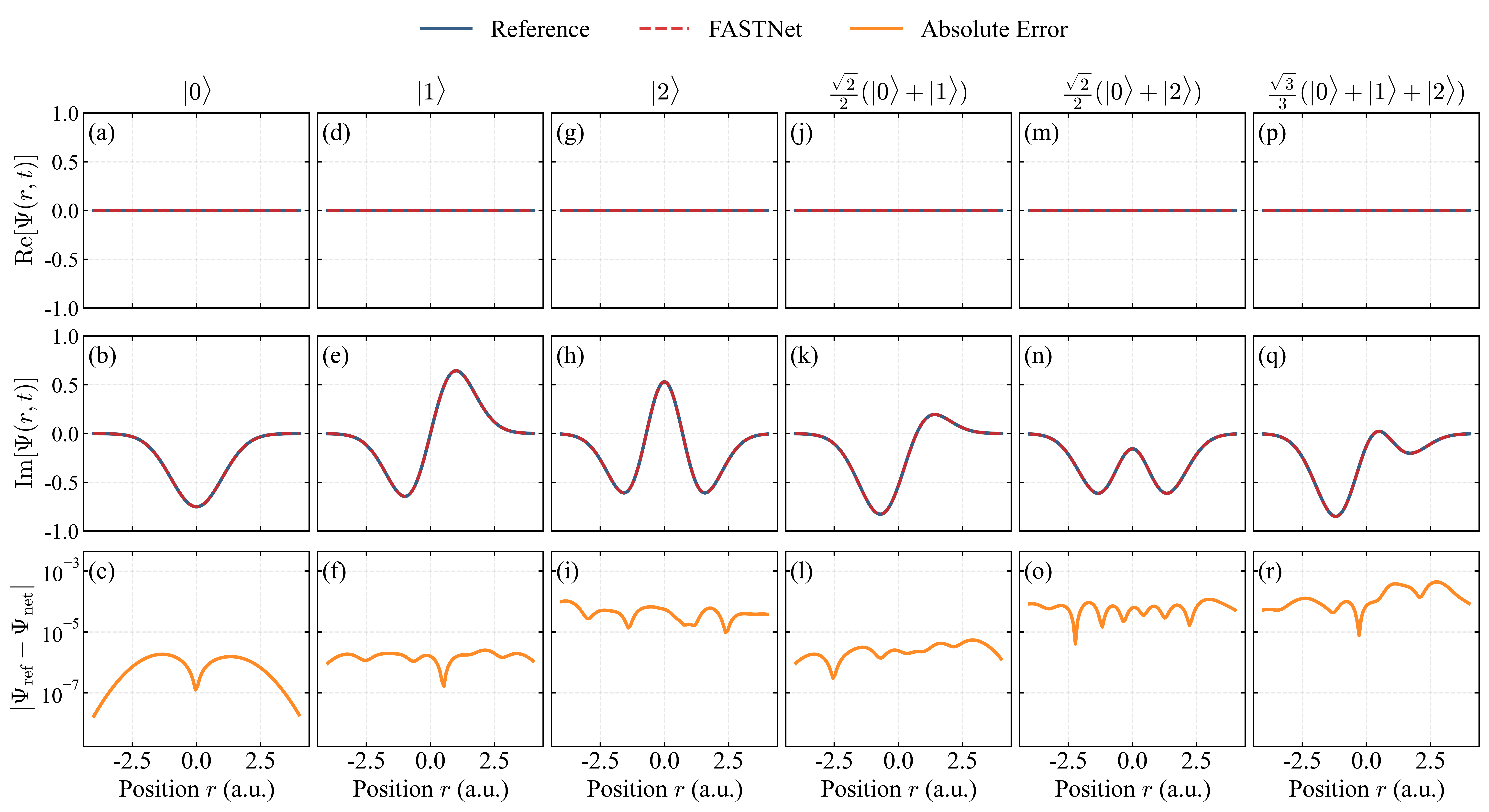}
    \caption{
      Real and imaginary parts, as well as absolute errors, of the 1D harmonic oscillator wavefunctions at $t=\pi$ for various initial states.
      Blue solid lines denote analytical solutions, red dashed lines show FASTNet predictions, and orange lines indicate the absolute error $|\Psi_{\mathrm{ref}} - \Psi_{\mathrm{net}}|$.
      FASTNet accurately reproduces both amplitude and phase across all cases.
    }
    \label{fig:1d_oscillator}
\end{figure}

We test FASTNet on several representative initial conditions over $t\in[0,\pi]$, including individual eigenstates ($|0\rangle$, $|1\rangle$, $|2\rangle$) and superpositions:
$\frac{\sqrt{2}}{2}(|0\rangle + |1\rangle)$,  
$\frac{\sqrt{2}}{2}(|0\rangle + |2\rangle)$,  
$\frac{\sqrt{3}}{3}(|0\rangle + |1\rangle + |2\rangle)$.  
Figure~\ref{fig:1d_oscillator} presents the real and imaginary components of the predicted wavefunctions, together with their pointwise absolute errors, at $t=\pi$. 
Across all cases, FASTNet achieves a relative $L_2$ error below $2\times 10^{-4}$ (Table~\ref{tab:1d_harmonic_oscillator_error}), showing excellent agreement with analytical solutions.

The relative $L_2$ error, $\varepsilon_{L_2}^{\mathrm{rel}}$, is calculated over the entire spacetime domain via high-order Gauss--Legendre quadrature integration: 
\begin{equation}
  \varepsilon_{L_2}^{\mathrm{rel}}
  = \frac{\left(\int_0^T \int_{\mathbb{R}}\left|\Psi_{\text{pred}}(r, t)-\Psi_{\text{ref}}(r, t)\right|^2 \dif{r} \dif{t}\right)^{1 / 2}}{\left(\int_0^T \int_{\mathbb{R}}\left|\Psi_{\text{ref}}(r, t)\right|^2 \dif{r} \dif{t}\right)^{1 / 2}}
\end{equation}
These results confirm that FASTNet can reliably capture smooth, oscillatory quantum dynamics in a controlled setting.

\begin{table}[htbp]
    \centering
    \caption{
        Relative space-time $L_2$ error $\varepsilon_{L_2}^{\mathrm{rel}}$ for pure and mixed initial states in a 1D harmonic oscillator. 
        Mixed states are defined as: 
        $|\psi_{01}\rangle = \frac{\sqrt{2}}{2}(|0\rangle + |1\rangle)$, 
        $|\psi_{02}\rangle = \frac{\sqrt{2}}{2}(|0\rangle + |2\rangle)$, 
        and 
        $|\psi_{012}\rangle = \frac{\sqrt{3}}{3}(|0\rangle + |1\rangle + |2\rangle)$.
    }
    \label{tab:1d_harmonic_oscillator_error}
    \begin{tabular}{lccc|ccc}
        \toprule
        & \multicolumn{6}{c}{\textbf{Initial State}} \\
        \cmidrule(lr){2-7}
        & \multicolumn{3}{c}{\textit{Pure States}} 
        & \multicolumn{3}{c}{\textit{Mixed States}} \\
        \cmidrule(lr){2-4} \cmidrule(lr){5-7}
        \textbf{Error} 
        & $|0\rangle$ 
        & $|1\rangle$ 
        & $|2\rangle$ 
        & $|\psi_{01}\rangle$ 
        & $|\psi_{02}\rangle$ 
        & $|\psi_{012}\rangle$ \\
        \midrule
        $\varepsilon_{L_2}^{\mathrm{rel}}$
        & $1.75 \times 10^{-6}$ 
        & $5.01 \times 10^{-6}$ 
        & $5.36 \times 10^{-5}$ 
        & $5.71 \times 10^{-6}$ 
        & $5.11 \times 10^{-5}$ 
        & $1.66 \times 10^{-4}$ \\
        \bottomrule
    \end{tabular}
\end{table}

\subsection{Interacting Fermions in 1D Harmonic Trap}
\label{sec:1d-interaction}

To assess the capability of FASTNet on interacting many-body systems, we consider fermions confined in a one-dimensional harmonic trap with time-dependent confinement and pairwise quadratic interactions. The potential term of the Hamiltonian is
\begin{equation}
    V(\bm{r}, t) = \sum_{i=1}^N \left[ \frac{1}{2} \omega(t)^2 r_i^2 + \frac{g(t)}{2} \sum_{j < i} (r_i - r_j)^2 \right],
\end{equation}
where $\omega(t)$ denotes the time-dependent trap frequency, $g(t)$ is the interaction strength, and $r_i \in \mathbb{R}$ is the position of the $i$-th fermion. 
For specific initial conditions and protocols, including a sudden frequency quench, this model admits analytical solutions~\cite{nys2024ab, zaluska2000soluble, gritsev2010scaling}.

We study the case of a quench from $\omega_0 = 1.0$ to $\omega_f = 2.0$ at $t = 0$. 
With appropriate choice of $g(t)$, the system exhibits a collective breathing mode characterized by monopole oscillations with period $T = \frac{\pi}{\omega_f}$. 
The corresponding analytical (unnormalized) solution is:
\begin{equation}
    \Psi(\bm{r}, t) = e^{-iE_0 \tau(t)}\operatorname{det}\left[\mathcal{V}\left( \frac{\bm{r}}{L(t)} \right) \right] \exp\left( -\alpha(t) \sum_{i=1}^N r_i^2 - \beta(t) \left( \sum_{i=1}^N r_i \right)^2 \right),
\end{equation}
where $\mathcal{V}(\bm{r}) = \prod_{1 \leq i < j \leq N}(r_j - r_i)$ is the Vandermonde determinant, and $L(t)$, $\tau(t)$, $\alpha(t)$, $\beta(t)$ are time-dependent functions defined in the \hyperref[sec:1d-trapped-interacting-fermions]{appendix C}.

The quadratic interaction term introduces long-range coupling that grows rapidly as particles separate, causing the ground-state energy before the quench
\[
E_0 = \frac{1 + (N^2 - 1) \sqrt{1+N}}{2},
\]
to scale super-quadratically with particle number, approximately as 
$O(N^{5/2})$.
As a result, systems with even moderate particle number exhibit strong correlations and numerical instability, making general-purpose solvers prone to divergence.
To ensure a stable benchmark, we limit our tests to $N = 2$ and $N = 3$ fermions.
The initial interaction strength is chosen to be $g(0) = 1.0$.
The quench also drives rapid variations in $g(t)$, further challenging the learning task by amplifying interaction effects.

\begin{figure}
    \centering
    \includegraphics[width=\textwidth]{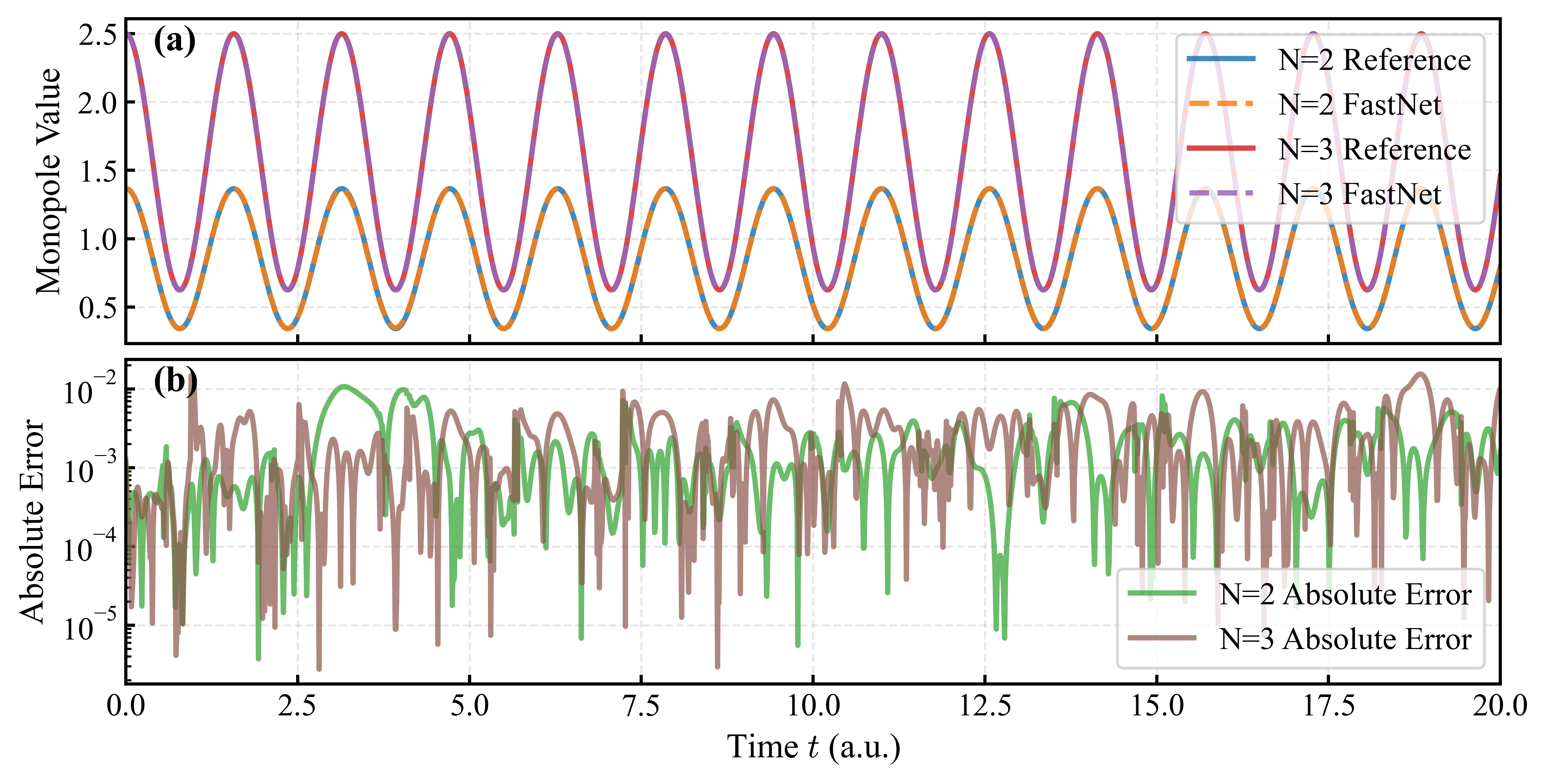}
    \caption{Results for $N = 2$ and $N = 3$ fermions. a) Monopole breathing mode; b) Absolute error between FASTNet and analytical solutions.}
    \label{fig:1d_interacting_fermions}
\end{figure}

Figure~\ref{fig:1d_interacting_fermions} presents the evolution of the monopole observable  $Q=\sum_i^N\left\langle r_i^2\right\rangle$.
In both cases, FASTNet accurately reproduces the breathing dynamics, with close agreement to the analytical reference. 
Notably, our architecture does not embed any knowledge of the analytical solution into the Ansatz—no explicit constraints on the envelope or phase are imposed. 
The only prior incorporated is the quadratic form ($p=2$) in the envelope term, directly reflecting the structure of the Hamiltonian.

\subsection{3D Hydrogen Orbital Evolution}

We expand FASTNet to three-dimensional systems, and apply it to the real-time evolution of a single electron in a hydrogen atom. 
In the absence of external fields and with the nucleus fixed at the origin, the system is governed by the time-dependent Schrödinger equation with Coulomb potential
\begin{equation}
    \hat{H} = -\frac{1}{2} \nabla^2 - \frac{1}{|\bm{r}|},
\end{equation}
where $\bm{r} \in \mathbb{R}^3$ denotes the position of the electron, and atomic units are used throughout. 
This system has exact analytical solutions, with stationary states of the form 
\begin{equation}
    \psi_{n\ell m}(\bm{r}) =\sqrt{\left(\frac{2}{n a}\right)^3 \frac{(n-\ell-1)!}{2 n(n+\ell)!}} e^{-r / n a}\left(\frac{2 r}{n a}\right)^{\ell}\left[L_{n-\ell-1}^{2 \ell+1}(2 r / n a)\right] Y_{\ell}^m(\theta, \phi),
\end{equation}
where $n$, $\ell$, and $m$ are the principal, angular momentum, and magnetic quantum numbers, respectively; $r$, $\theta$, and $\phi$ are the spherical coordinates of $\bm{r}$. 
$L_{n-\ell-1}^{2 \ell+1}$ represents the associated Laguerre polynomial, $Y_{\ell}^m$ is the spherical harmonic function. 
The energy of each eigenstate depends only on the principal quantum number $n$, given by $E_n = -\frac{1}{2n^2}$ for integer $n \geq 1$. 
Time evolution is then given by $\Psi(\bm{r}, t) = \sum_{n\ell m} c_{n\ell m} \psi_{n\ell m}(\bm{r}) e^{-i E_n t}$ for an initial state $\Psi(\bm{r}, 0) = \sum_{n\ell m} c_{n\ell m} \psi_{n\ell m}(\bm{r})$.

\begin{table}[htbp]
    \centering
    \caption{Relative space-time $L_2$ error $\varepsilon_{L_2}^{\mathrm{rel}}$ for pure and mixed hydrogenic initial states of the 3D hydrogen atom over $t\in[0,\pi]$.}
    \label{tab:3d_hydrogen_error}
    \begin{tabular}{lccccc}
        \toprule
        \textbf{Method} 
        & \multicolumn{5}{c}{\textbf{Initial State}} \\
        \cmidrule(lr){2-6}
        & \multicolumn{4}{c}{\textit{Pure States}} 
        & \\
        \cmidrule(lr){2-5}
        & $|1s\rangle$ 
        & $|2s\rangle$ 
        & $|2p_z\rangle$ 
        & $|3s\rangle$ 
        & \\
        \midrule
        FCI/aug-cc-pVQZ 
        & $ 2.90 \times 10^{-3}$
        & $ 8.16 \times 10^{-2}$
        & $ 4.66 \times 10^{-1}$
        & $ 1.13 \times 10^{0}$ 
        & \\
        FCI/d-aug-cc-pVQZ 
        & $ 2.89 \times 10^{-3}$
        & $ 2.18 \times 10^{-2}$
        & $ 2.88 \times 10^{-2}$
        & $ 1.07 \times 10^{-1}$ 
        & \\          
        FastNet 
        & $ 8.33 \times 10^{-5}$
        & $ 1.21 \times 10^{-4}$
        & $ 1.23 \times 10^{-4}$
        & $ 2.33 \times 10^{-4}$ 
        & \\ 
        \midrule
        & \multicolumn{4}{c}{\textit{Mixed States}} 
        & \\
        \cmidrule(lr){2-5}
        & $|1s2p_z\rangle$ 
        & $|2p_x 2p_z\rangle$ 
        & $|1s2s3s\rangle$ 
        & $|2s2p_z 3d_{z^2}\rangle$ \\     
        \midrule
        FCI/aug-cc-pVQZ 
        & $ 3.30 \times 10^{-1}$
        & $ 4.66 \times 10^{-1}$
        & $ 6.01 \times 10^{-1}$ 
        & $ 7.99 \times 10^{-1}$ \\
        FCI/d-aug-cc-pVQZ 
        & $ 2.05 \times 10^{-2}$
        & $ 2.88 \times 10^{-2}$
        & $ 7.20 \times 10^{-2}$
        & $ 5.09 \times 10^{-1}$ \\          
        FastNet 
        & $ 3.82 \times 10^{-4}$ 
        & $ 1.83 \times 10^{-4}$ 
        & $ 2.78 \times 10^{-4}$ 
        & $ 1.17 \times 10^{-3}$ \\ 
        \bottomrule
    \end{tabular}
\end{table}

Gaussian-type orbital (GTO) basis sets are ubiquitous in quantum chemistry, their smooth, well-behaved functional form makes them particularly suitable for representing a broad range of atomic and molecular orbitals. 
However, the Gaussian radial decay poorly matches the long-range exponential tails and nodal structure of highly excited (Rydberg-like) states. 
We quantify this by propagating the TDSE with full configuration interaction (FCI) in two increasingly diffuse bases aug-cc-pVQZ and d-aug-cc-pVQZ and by comparing with FASTNet, using the exact analytical evolution as reference. 
We test the pure states $|1s\rangle$, $|2s\rangle$, $|2p_z\rangle$, and $|3s\rangle$; here $|2p_z\rangle$ denotes the real Cartesian hydrogenic orbital, i.e., a geometrically oriented real combination of the $|n\ell m\rangle$ eigenstates (with $s,p$ corresponding to $\ell=0,1$ and $p_z$ corresponding to $m=0$, while $p_x,p_y$ arise from real mixes of $m=\pm1$).
The relative space-time $L_2$ errors (integrated over $t\in[0,\pi]$) are reported in Table~\ref{tab:3d_hydrogen_error}. 
With aug-cc-pVQZ(46 bases), the relative $L_2$ error is already small for $|1s\rangle$ but grows rapidly with excited states $|2s\rangle$ and $|2p_z\rangle$, and becomes catastrophic for the highly diffuse state $|3s\rangle$. 
Adding a second set of diffuse functions (d-aug-cc-pVQZ, 62 bases) alleviates but does not remove the mismatch, still cannot adequately capture excited states.
These trends underscore that, even with augmentation, standard GTO families are not ideally suited for highly excited hydrogenic states.
By working directly in real space, FASTNet sidesteps this basis-family limitation and achieves errors in the $10^{-4}$ range across all tested pure states, roughly two orders of magnitude lower than d-aug-cc-pVQZ and, for $|3s\rangle$, nearly four orders lower than aug-cc-pVQZ. 
Together, these results indicate that FASTNet faithfully represents and propagates highly excited hydrogenic states in a three-dimensional Coulomb potential. The architecture explicitly accommodates long-range behavior, suggesting potential applicability to Rydberg-like manifolds; a systematic assessment on genuine Rydberg atoms is left for future work.

To provide a stringent benchmark beyond pure eigenstates, we focus on nontrivial superpositions of hydrogenic orbitals, which exhibit richer spatial structure and more intricate phase evolution. 
The initial states considered include $|1s\,2p_z\rangle$, $|2p_x\,2p_z\rangle$, $|1s\,2s\,3s\rangle$, and $|2s\,2p_z\,3d_{z^2}\rangle$.
Using the real Cartesian hydrogenic representation, $d$ states correspond to $\ell=2$, with $d_{z^2}$ representing the $m=0$ orbital.
Here $|1s\,2p_z\rangle = \frac{1}{\sqrt{2}}(\psi_{100}+\psi_{210})$, which is chosen as a physically motivated state: it can be accessed by coherent laser excitation with a linearly polarized field along the $z$ axis.
Other three states, $|2p_x\,2p_z\rangle = \frac{1}{\sqrt{3}}(\psi_{210}-\psi_{211}+\psi_{21-1})$, $|1s\,2s\,3s\rangle = \frac{1}{\sqrt{3}}(\psi_{100}+\psi_{200}+\psi_{300})$, and $|2s\,2p_z\,3d_{z^2}\rangle = \frac{1}{\sqrt{3}}(\psi_{200}+\psi_{210}+\psi_{320})$, are constructed to probe the representation of mixed states with varying angular momentum and radial characteristics.

We evaluate the performance of FASTNet against FCI calculations using aug-cc-pVQZ and d-aug-cc-pVQZ basis sets, quantifying accuracy through the relative $L_2$ error integrated over the space-time domain.
Simulations span the temporal interval $t\in[0,\pi]$, with results presented in Table~\ref{tab:3d_hydrogen_error}.
Consistent with our findings in the previous part, finite GTO basis sets incur large representation errors for highly excited and diffuse states, even with the d-aug-cc-pVQZ basis, and demonstrate particularly poor performance for mixed states.
In contrast, FASTNet achieves uniformly small relative $L_2$ errors across all tested superpositions, typically about two orders of magnitude lower than the errors of FCI/d-aug-cc-pVQZ.
These results demonstrate that FASTNet reliably captures real-time quantum dynamics in a three-dimensional Coulomb potential, with accuracy preserved even for complex mixed states involving multiple orbitals.

\subsection{Hydrogen Atom in an Elliptically Polarized Laser Field}

Building upon the validation of the 3D hydrogen atom in field-free conditions, we now challenge the framework with an explicitly time-dependent, external field. 
Specifically, we consider the real-time dynamics of a hydrogen atom subjected to an elliptically polarized laser field. 
The system evolves over a temporal interval of $t \in [0, 30]$ a.u., encompassing both the pulse duration and the subsequent field-free evolution.

The dynamics are governed by the time-dependent Schr\"odinger equation with the overall Hamiltonian in the length gauge:
\begin{equation}
    \hat{H}(t) = -\frac{1}{2}\nabla^2 - \frac{1}{|\mathbf{r}|} + \mathbf{r} \cdot \mathbf{E}(t),
\end{equation}
where $\mathbf{r} = (x, y, z)$ denotes the spatial coordinate of the electron. 
The atom is driven by a laser pulse characterized by the electric field $\mathbf{E}(t)$ confined to the $x$-$y$ plane:
\begin{equation}
    \mathbf{E}(t) = \frac{\mathcal{E}_{\max}}{\sqrt{1+\epsilon^2}} f(t) \left[\sin(\omega t)\mathbf{\hat{x}} + \epsilon \cos(\omega t)\mathbf{\hat{y}}\right].
\end{equation}
The pulse envelope $f(t)$ takes a $\sin^2$ form:
\begin{equation}
    f(t) = \begin{cases} 
        \sin^2\left(\frac{\pi t}{nT}\right), & 0 \le t \le nT \\ 
        0, & \text{otherwise} 
    \end{cases},
\end{equation}
where $T = 2\pi/\omega$ is the optical period. 
For this simulation, we set the peak field strength to $\mathcal{E}_{\max} = 0.05$ a.u., the carrier frequency to $\omega = 1.0$ a.u., the ellipticity to $\epsilon = 0.8$, and the number of optical cycles to $n = 2$.

The introduction of this elliptically polarized field explicitly breaks the spherical symmetry of the Coulomb potential, thereby precluding an exact analytical solution. 
However, because this is a single-electron system, the computational curse of dimensionality is avoided. 
This dimensional tractability allows us to employ a high accuracy grid-based numerical solver to solve the real-space time-dependent Schr\"odinger equation directly~\cite{zhang2025spin,ma2025complete}, establishing a strict ground truth for rigorous benchmarking.

\begin{figure}
    \centering
    \includegraphics[width=1.0\textwidth]{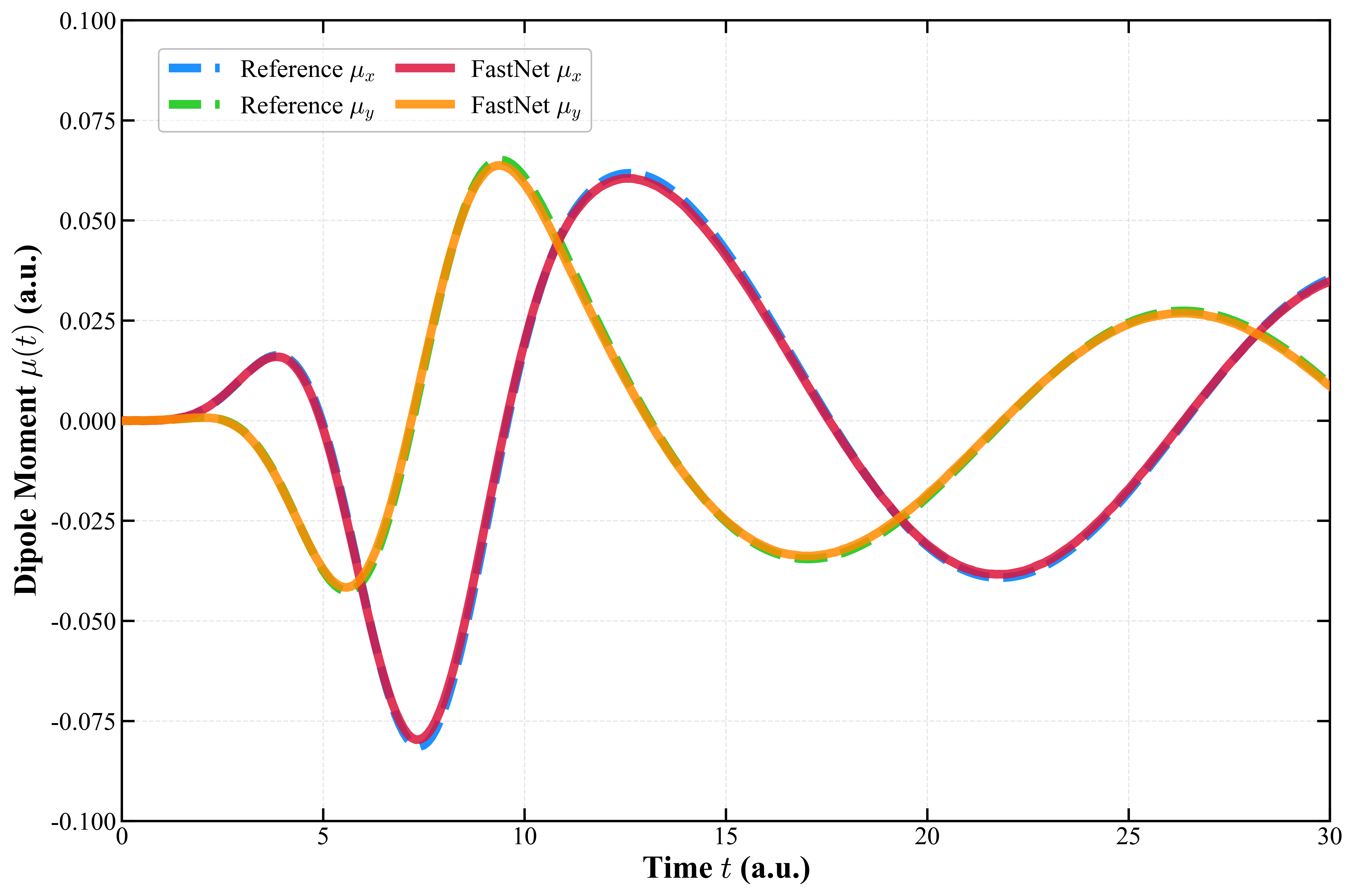}
    \caption{Time evolution of the induced dipole moment components, $\mu_x(t)$ and $\mu_y(t)$, for a hydrogen atom driven by an elliptically polarized laser field.}
    \label{fig:dipole_hep}
\end{figure}

Figure~\ref{fig:dipole_hep} illustrates the time evolution of the induced dipole moment components, $\mu_x(t) = -\langle x \rangle$ and $\mu_y(t) = -\langle y \rangle$, under the elliptically polarized field. 
The FASTNet predictions exhibit overall strong quantitative agreement with the numerical reference across the entire 30 a.u. time interval. 
The network successfully captures the primary field-induced polarization, albeit with a slight underestimation of the peak oscillation amplitudes. 
This minor deviation primarily stems from the strict spatial constraint imposed by the exponentially decaying envelope. 
While essential for stabilizing bound-state dynamics, this localized \textit{ansatz} inherently restricts the network's flexibility to fully represent the extreme spatial distension of the wavefunction under strong laser fields.

\subsection{Molecular Dynamics in Linearly Polarized Laser Fields}

As a demanding test case, we simulate the real-time dipole response of a stretched hydrogen molecule (H$_2$) driven by linearly polarized laser pulses.
Scaling up from single-electron 3D systems to multi-electron dynamics introduces the curse of dimensionality, making direct real-space grid computations exceedingly demanding.
Consequently, we benchmark FASTNet against basis-dependent quantum chemistry methods, specifically Time-Dependent Hartree-Fock (TD-HF) and Full Configuration Interaction (FCI). 
The dynamics obey the TDSE with the Hamiltonian
\begin{equation}
H(t)=\sum_{i=1}^{N}\!\left(-\frac{1}{2}\nabla_i^2\right)
+\sum_{i<j}\frac{1}{|\bm r_i-\bm r_j|}
-\sum_{i=1}^{N}\sum_{a=1}^{N_A}\frac{Z_a}{|\bm r_i-\bm R_a|}
-E(t)\sum_{i=1}^{N} x_i,
\end{equation}
i.e., the length-gauge coupling to a spatially uniform field along $x$. 
For H$_2$ we set $N=2$, $N_A=2$, $Z_1=Z_2=1$, and place the nuclei on the $x$ axis at $\bm R_1=(-1.393038,0,0)$ and $\bm R_2=(1.393038,0,0)$ a.u., corresponding to a bond length of $2.786076$ a.u. (twice the equilibrium value).
The laser electric field has peak strength $\mathcal{E}_{\max}$ a.u. and time profile
\begin{equation}
E(t)=\mathcal{E}_{\max}\, f(t)\sin(\omega t),\qquad
f(t)=
\begin{cases}
t/T, & 0\le t<T,\\
1, & T\le t<2T,\\
3-t/T, & 2T\le t<3T,\\
0, & \text{otherwise},
\end{cases}
\end{equation}
with $T=2\pi/\omega$, $\omega=0.1$. 
Details regarding the computational cost and hardware specifications for these simulations are provided in Appendix~\ref{sec:computational-resources-and-cost}.
We deliberately consider this stretched, non-equilibrium geometry where mean-field approximations (like TD-HF) are known to struggle with strong correlation and bond dissociation, providing a stringent test for tracking correlation-driven coherence.

To systematically evaluate the performance of FASTNet, we dissect the problem into two regimes: a weak-field regime to isolate and verify the network's capacity to represent dynamic electron correlation, and a strong-field regime to benchmark its stability and expressivity under highly non-equilibrium conditions.

\begin{figure}
    \centering
    \includegraphics[width=1.0\textwidth]{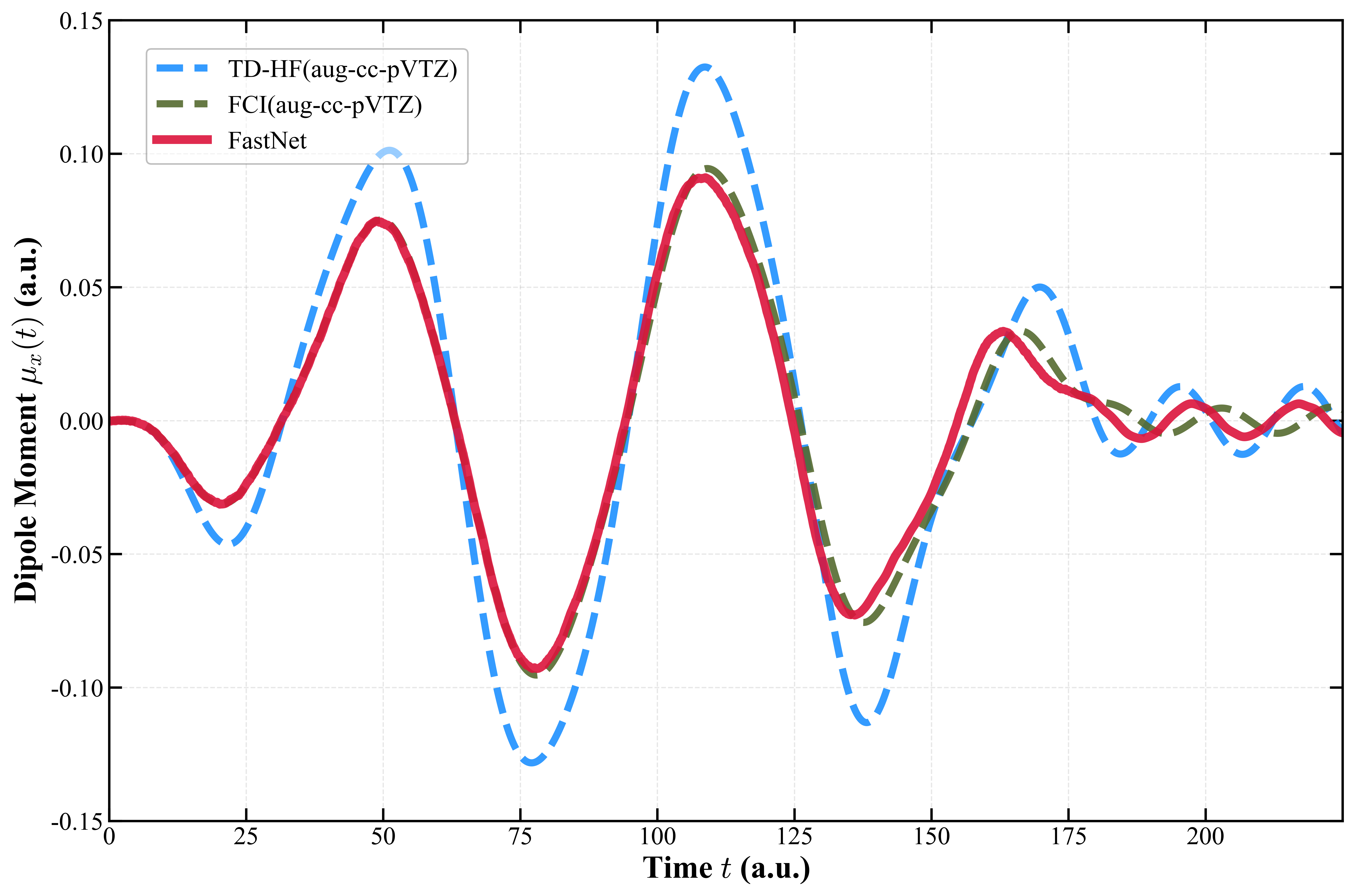}
    \caption{Time evolution of the induced $x$-component of the electronic dipole $\mu_x(t)$ for stretched H$_2$ driven by the weak linearly polarized laser field.}
    \label{fig:dipole_h2w}
\end{figure}

We first simulate the dynamics under a weak external perturbation ($\mathcal{E}_{\max} = 0.005$ a.u.), and the system is initialized in ground state and evolved for a total simulation time extending beyond three laser periods. 
In this regime, field-induced ionization is negligible, and the wavefunction evolution is dominated by bound-state transitions. 
Under these conditions, both FCI and TD-HF can achieve their respective basis-set convergence limits when employing a large, augmented Gaussian basis set (aug-cc-pVTZ). 
Figure~\ref{fig:dipole_h2w} illustrates the induced $x$-component of the dipole moment $\mu_x(t)$. 
As shown, while the converged TD-HF(aug-cc-pVTZ) captures the basic oscillatory response, it exhibits noticeable phase and amplitude deviations from the FCI reference due to its mean-field nature. 
FASTNet, however, closely tracks the FCI(aug-cc-pVTZ) dynamics throughout the entire simulation. 
This strong agreement verifies that our explicitly antisymmetric spatiotemporal Ansatz successfully resolves the correlation-driven coherent dynamics.

Moving beyond the weak-field verification, we now challenge FASTNet deep in the non-perturbative regime. 
At a peak field strength of $\mathcal{E}_{\max} = 0.07$ a.u., the laser-matter interaction is highly nonlinear. 
Under such extreme conditions, dynamics such as tunnel ionization and the population of diffuse Rydberg or continuum states become prominent. 
Consequently, spatially localized Gaussian basis sets --- even extensively augmented ones --- struggle to converge efficiently. 
Therefore, rather than seeking an absolute physical ground truth, we adopt a strict cross-method benchmarking protocol. 
To establish a rigorous comparison with the time-dependent variational Monte Carlo (tVMC) approach, we explicitly adopt the exact benchmark setup and basis sets (STO-3G and cc-pVDZ) established by Nys et al. in Ref.~\cite{nys2024ab}.
Accordingly, the reference trajectories for TD-HF, FCI, and tVMC presented in this regime are obtained from their work to ensure a comparison on equal footing.

\begin{figure}
    \centering
    \includegraphics[width=1.0\textwidth]{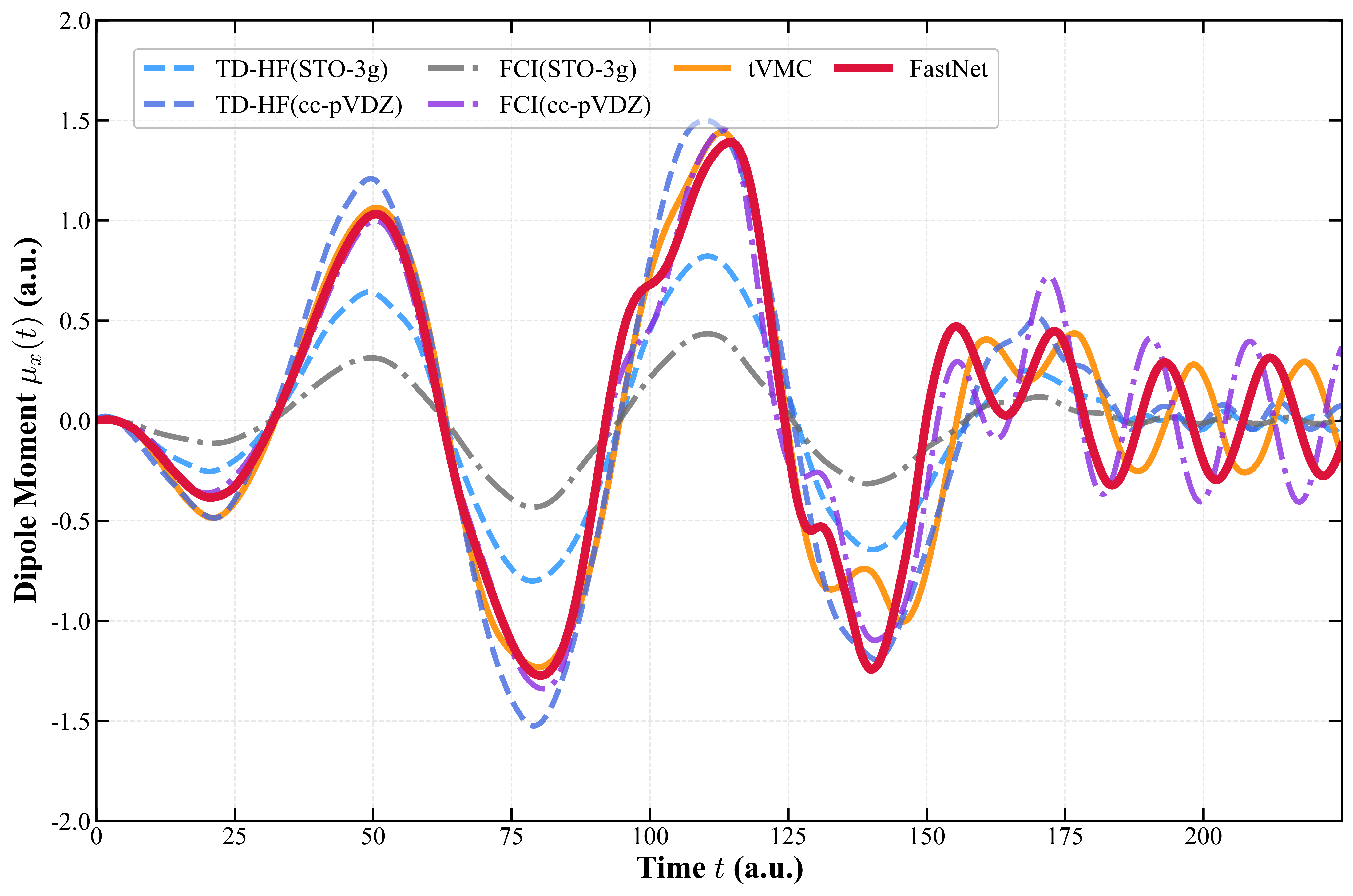}
    \caption{Time evolution of the induced $x$-component of the electronic dipole $\mu_x(t)$ for stretched H$_2$ driven by the strong linearly polarized laser field.}
    \label{fig:dipole_h2s}
\end{figure}

Figure~\ref{fig:dipole_h2s} compares the induced $x$-component of the dipole moment $\mu_x(t)$, obtained with different methods.
For basis-dependent methods, the error budgets differ.
Neglecting long-time integrator drift, FCI is exact within a finite orbital basis and thus primarily limited by basis-set incompleteness, whereas TD-HF suffers from both basis error and a systematic mean-field bias due to the neglect of electron correlation. 
Consistent with enhanced polarizability, enlarging the basis from STO-3G to cc-pVDZ increases the overall oscillation amplitude for each method. 
During the pulse, TD-HF yields larger amplitudes than the correlated method FCI. 
After the laser field is switched off, FCI(cc-pVDZ) exhibits a damped, field-free oscillation, while TD-HF shows little to no residual oscillation, reflecting its inability to sustain the correlation-driven coherent dynamics.

Just as finite GTO bases fail to capture full ionization, FASTNet, in its current formulation, is not exempt from the challenges of continuum dynamics. 
The exponentially decaying envelope functions embedded in our Ansatz inherently enforce spatial localization, which artificially restricts extensive ionization flux and acts as a boundary constraint. 
Therefore, the deviations observed in peak heights and late-time side peaks among FASTNet, tVMC, and FCI(cc-pVDZ) reflect the different structural biases of each restricted Ansatz when pushed toward the continuum limit.
Nevertheless, within this benchmarking protocol, the trajectory of FASTNet is highly instructive. 
The neural-network approach tVMC closely tracks FCI(cc-pVDZ) in both the driven and field-free windows. 
FASTNet similarly exhibits excellent agreement with the FCI(cc-pVDZ) baseline within the bound-state manifold, recovering the peak heights and the correlated post-pulse oscillation. 
This indicates that, despite constraints on modeling full ionization, our spacetime optimization successfully captures the correlation-driven coherent dynamics that are fundamentally absent in mean-field descriptions.

In summary, while the current formulation's localized envelope imposes physical reservations regarding the exact representation of extensive ionization flux in extreme fields, these results demonstrate the robustness of our approach. 
By effectively capturing the time-dependent build-up and decay of multi-electron correlations, FASTNet successfully bridges the critical gap between mean-field approximations and fully correlated \textit{ab initio} methods, providing an expressive framework for real-time quantum dynamics.

\section{Conclusion}

In this work, we introduced \textbf{FASTNet}, a neural-network framework for solving the TDSE in continuous real space.
By treating time as an explicit network input and formulating the TDSE as a global spacetime optimization problem, FASTNet provides a complementary route to traditional stepwise propagation methods.
The antisymmetric neural ansatz ensures fermionic antisymmetry, while the pretraining strategy enables accurate long-time simulations.

We validated FASTNet on a hierarchy of benchmarks, from a 1D harmonic oscillator and interacting fermions in a 1D harmonic trap to the 3D hydrogen atom and a laser-driven H$_2$ molecule. 
Across these increasingly complex systems, FASTNet consistently reproduced analytic or high-level correlated reference results in perturbative and bound-state regimes. 
Furthermore, it successfully captured correlation-induced interference effects beyond mean-field descriptions, although discrepancies remain in highly non-linear regimes where continuum transitions become prominent.

Looking ahead, we will improve training stability via adaptive spacetime sampling, mitigate error accumulation in long-time propagation, and extend the framework to larger, strongly correlated, and open quantum systems.
Regarding scalability to larger systems, while the network evaluation scales polynomially, practical bottlenecks are expected to arise primarily from MCMC sampling efficiency in high-dimensional configuration space and the increasing complexity of the optimization.
Addressing these challenges will be a prerequisite for extending the framework to larger systems.
Furthermore, the current framework employs a fixed spin assignment, which effectively restricts the scope to non-relativistic regimes. 
Extending the method to describe spin-flip dynamics and spin-orbit coupling remains an important goal for generalizing the ansatz.
On the application side, we will also evaluate FASTNet for strong-field ultrafast dynamics. 
Its global-in-time, real-space, antisymmetric design targets multi-electron correlation beyond mean-field and appears well suited to nonsequential double ionization, multielectron recombination in high-harmonic generation, and photoionization time delays. 
We will investigate strategies to extend the ansatz to capture continuum states in ionization modeling, incorporate gauge-consistent laser-matter coupling, and explore extensions beyond the Born-Oppenheimer approximation, with the broader goal of developing a neural-network-based workflow for quantitative ultrafast dynamics.
Taken together, these efforts position FASTNet as a scalable, expressive, and generalizable tool for real-time quantum dynamics, complementing existing step-by-step approaches and broadening the scope of \textit{ab initio} simulations in quantum chemistry and materials science.

\section*{Acknowledgements}

We are grateful to Liangyou Peng, Ji Chen and Weizhong Fu for fruitful discussions.
The work of Enze Hou and Han Wang is supported by the National Key R\&D Program of China (Grant No.~2022YFA1004300) and the National Natural Science Foundation of China (Grants No.~12525113 and No.~12561160120).

\bibliographystyle{unsrtnat}
\bibliography{main}

\newpage
\appendix
\newpage

\section{Network Evaluation Algorithm and Scaling}

The wavefunction evaluation procedure formalizes as:

\begin{algorithm}[H]
\caption{FASTNet evaluation}
\label{alg:ferminet}
\begin{algorithmic}[1]
    \Require Walker configuration $\{\bm{r}_1^\uparrow, \cdots, \bm{r}_n^\uparrow, \bm{r}_1^\downarrow, \cdots, \bm{r}_{n_l}^\downarrow\}$, Nuclear positions $\{\bm{R}_I\}$, Time $t$
    
    \For{each electron $i$, $\alpha$}
        \State $\displaystyle
            \bm{h}_{i}^{\ell\alpha} \leftarrow \text{concatenate} \left( \bm{r}_{i}^{\alpha} - \bm{R}_I,\ |\bm{r}_{i}^{\alpha} - \bm{R}_I|,\ t \quad \forall I \right)
        $
        \State $\displaystyle
            \bm{h}_{ij}^{\ell\alpha\beta} \leftarrow \text{concatenate} \left( \bm{r}_{i}^{\alpha} - \bm{r}_{j}^{\beta},\ |\bm{r}_{i}^{\alpha} - \bm{r}_{j}^{\beta}|,\ t \quad \forall j, \beta \right)
        $
    \EndFor
    
    \For{each layer $\ell \in \{0, L-1\}$}
        \State $\bm{g}^{\ell\uparrow} \leftarrow \frac{1}{n^{\uparrow}} \sum_{i}^{n^{\uparrow}} \bm{h}_{i}^{\ell\uparrow}$
        \State $\bm{g}^{\ell\downarrow} \leftarrow \frac{1}{n^{\downarrow}} \sum_{i}^{n^{\downarrow}} \bm{h}_{i}^{\ell\downarrow}$
        
        \For{each electron $i$, $\alpha$}
            \State $\bm{g}_{i}^{\ell\alpha\uparrow} \leftarrow \frac{1}{n^{\uparrow}} \sum_{j}^{n^{\uparrow}} \bm{h}_{ij}^{\ell\alpha\uparrow}$
            \State $\bm{g}_{i}^{\ell\alpha\downarrow} \leftarrow \frac{1}{n^{\downarrow}} \sum_{j}^{n^{\downarrow}} \bm{h}_{ij}^{\ell\alpha\downarrow}$
            \State $\bm{f}_{i}^{\ell\alpha} \leftarrow \text{concatenate} \left( \bm{h}_{i}^{\ell\alpha},\ \bm{g}^{\ell\uparrow},\ \bm{g}^{\ell\downarrow},\ \bm{g}_{i}^{\ell\alpha\uparrow},\ \bm{g}_{i}^{\ell\alpha\downarrow} \right)$
            \State $\bm{h}_{i}^{\ell+1\alpha} \leftarrow \tanh\left( \text{matmul}(\bm{V}^{l}, \bm{f}_{i}^{\ell\alpha}) + \bm{b}^{l} \right) + \bm{h}_{i}^{\ell\alpha}$
            \State $\bm{h}_{ij}^{\ell+1\alpha\beta} \leftarrow \tanh\left( \text{matmul}(\bm{W}^{l}, \bm{h}_{ij}^{\ell\alpha\beta}) + \bm{c}^{l} \right) + \bm{h}_{ij}^{\ell\alpha\beta}$
        \EndFor
    \EndFor
    
    \For{each determinant $k$}
        \For{each orbital $i$}
            \For{each electron $j$, $\alpha$}
                \State $e \leftarrow \text{envelope} \left( \bm{r}_{j}^{\alpha},\ \{\bm{r}_{i}^{\alpha} - \bm{R}_I\} \right)$
                \State $J \leftarrow \text{phase factor} \left(\exp\left[\mathrm{i}\, S^{i k}_\theta(\bm{h}_j^{L\alpha}, t)\right] \right)$
                \State $\phi_{i} (\bm{r}_{j}^{\alpha},\ t;\ \{\bm{r}_{/j}^{\alpha}\};\ \{\bm{r}^{\bar{\alpha}}\}) = \left( \text{dot} \left( \bm{w}_{i}^{k\alpha},\ \bm{h}_{j}^{L\alpha} \right) + g_{i}^{k\alpha} \right) e J$
            \EndFor
        \EndFor
        \State $D^{k\uparrow} \leftarrow \det \left[ \phi_{i}^{k\uparrow} (\bm{r}_{j}^{\uparrow},\ t;\ \{\bm{r}_{/j}^{\uparrow}\};\ \{\bm{r}^{\downarrow}\}) \right]$
        \State $D^{k\downarrow} \leftarrow \det \left[ \phi_{i}^{k\downarrow} (\bm{r}_{j}^{\downarrow},\ t;\ \{\bm{r}_{/j}^{\downarrow}\};\ \{\bm{r}^{\uparrow}\}) \right]$
    \EndFor
    
    \State $\psi \leftarrow \sum_{k} \omega_k D^{k\uparrow} D^{k\downarrow}$
\end{algorithmic}
\end{algorithm}

The computational complexity of FASTNet per training step is theoretically dominated by the evaluation of Slater determinants.
In terms of the network architecture, one-electron stream evaluating scales as $\mathcal{O}(N)$, the evaluation of two-electron stream, envelope, phase factor, and base amplitude scales as $\mathcal{O}(N^2)$, and determinant evaluating scales as $\mathcal{O}(N^3)$. 
Consequently, the total cost of evaluating the wavefunction scales as $\mathcal{O}(N^3)$, governed by the determinant calculation.
Furthermore, by employing the Forward Laplacian method~\cite{li2024computational}, the computation of gradients and the kinetic energy achieves the same asymptotic complexity as the function evaluation itself.
Therefore, the overall algorithmic scaling for a single training step is $\mathcal{O}(N^3)$.

\section{Loss and Gradient}
\label{sec:loss-gradient}

For simplicity, we omit the network parameters $\eta$ in the wavefunction $\Psi_\eta$ throughout this section and simply denote it as $\Psi$.

For the initial loss $\mathcal{L}_{I}$, sampling is performed according to the normalized square of the initial wavefunction, which is a probability density independent of network parameters, allowing direct autodiff gradient computation on the discretized form. 
For the residual loss $\mathcal{L}_{R}$, the sampling function is the normalized square of the parameterized wavefunction, making direct differentiation of the discretized form infeasible. 
Instead, we need to compute an unbiased estimate of its gradient under sampling. 
Denoting the network parameters as $\eta$, we have:
\begin{equation}
    \begin{aligned}
    \nabla_\eta \mathcal{L}_{R} & =\nabla_\eta \int_{\Omega}\int_0^T \dfrac{1}{T}\dfrac{1}{\langle\Psi|\Psi\rangle} \left| i\dfrac{\partial\Psi}{\partial t}(\bm{r},t) - E_L(\bm{r},t) \Psi(\bm{r},t) \right|^2 \dif{\bm{r}}\dif{t} \\
    \nabla_\eta \mathcal{L}_{R} & =\nabla_\eta \int_{\Omega}\int_0^T \dfrac{1}{T}\dfrac{\Psi^*\Psi}{\langle\Psi|\Psi\rangle} \left| i\dfrac{\partial\log\Psi}{\partial t}(\bm{r},t) - E_L(\bm{r},t) \right|^2 \dif{\bm{r}}\dif{t} \\
    & = \int_{\Omega}\int_0^T \dfrac{1}{T}(\nabla_\eta\dfrac{\Psi^*\Psi}{\langle\Psi|\Psi\rangle})\left| i\dfrac{\partial\log\Psi}{\partial t}(\bm{r},t)-E_L(\bm{r},t) \right|^2 \dif{\bm{r}}\dif{t} \\
    & \quad + \int_{\Omega}\int_0^T \dfrac{1}{T}\dfrac{\Psi^*\Psi}{\langle\Psi|\Psi\rangle}(\nabla_\eta\left| i\dfrac{\partial\log\Psi}{\partial t}(\bm{r},t) - E_L(\bm{r},t) \right|^2) \dif{\bm{r}}\dif{t} \\
    & = \int_{\Omega}\int_0^T \dfrac{1}{T}\dfrac{\Psi^*\Psi}{\langle\Psi|\Psi\rangle}(\dfrac{\nabla_\eta\Psi^*}{\Psi^*} + \dfrac{\nabla_\eta\Psi}{\Psi})\left| i\dfrac{\partial\log\Psi}{\partial t}(\bm{r},t) - E_L(\bm{r},t) \right|^2 \dif{\bm{r}}\dif{t}\\
    & \quad - \int_{\Omega}\int_0^T \dfrac{1}{T}\dfrac{\Psi^*\Psi}{\langle\Psi|\Psi\rangle} \dfrac{\langle\nabla_\eta\Psi|\Psi\rangle+\langle\Psi|\nabla_\eta\Psi\rangle}{\langle\Psi|\Psi\rangle}\left| i\dfrac{\partial\log\Psi}{\partial t}(\bm{r},t) - E_L(\bm{r},t) \right|^2 \dif{\bm{r}}\dif{t}\\
    & \quad + \int_{\Omega}\int_0^T \dfrac{1}{T}\dfrac{\Psi^*\Psi}{\langle\Psi|\Psi\rangle}(\nabla_\eta\left| i\dfrac{\partial\log\Psi}{\partial t}(\bm{r},t) - E_L(\bm{r},t)\right|^2) \dif{\bm{r}}\dif{t} 
    \end{aligned}
\end{equation}

Note that in the second term above, $\dfrac{\langle\nabla_\eta\Psi|\Psi\rangle+\langle\Psi|\nabla_\eta\Psi\rangle}{\langle\Psi|\Psi\rangle}$ no longer depends on $\bm{r}$ after integration over space (via the bracket notation). 
This allows us to compute it as an expectation outside the spatial integral. 
By discretizing time and sampling the time domain $[0,T]$ into $N_t$ equal intervals, with $N_r$ sampling points in each interval, giving a total of $N_R=N_t \times N_r$ unsupervised residual sampling points, we can obtain an unbiased estimate of $\mathcal{L}_{res}$ as
\begin{equation}
    \begin{aligned}
    \nabla_\eta \mathcal{L}_{R} & =\nabla_\eta \dfrac{1}{N_t}\sum_{k=1}^{N_t}\int_{\Omega}\dfrac{\Psi^*\Psi}{\langle\Psi|\Psi\rangle}(\bm{r},t^k)  \left| i\dfrac{\partial\log\Psi}{\partial t}(\bm{r},t^k) - E_L(\bm{r},t^k) \right|^2 \dif{\bm{r}} \\
    & = \dfrac{1}{N_t}\sum_{k=1}^{N_t}\int_{\Omega}\dfrac{\Psi^*\Psi}{\langle\Psi|\Psi\rangle}(\bm{r},t^k)(\dfrac{\nabla_\eta\Psi^*}{\Psi^*} + \dfrac{\nabla_\eta\Psi}{\Psi})(\bm{r},t^k)\left| i\dfrac{\partial\log\Psi}{\partial t}(\bm{r},t^k) - E_L(\bm{r},t^k) \right|^2 \dif{\bm{r}}\\
    & \quad - \dfrac{1}{N_t}\sum_{k=1}^{N_t}\int_{\Omega}\dfrac{\Psi^*\Psi}{\langle\Psi|\Psi\rangle}(\bm{r},t^k) \dfrac{\langle\nabla_\eta\Psi|\Psi\rangle+\langle\Psi|\nabla_\eta\Psi\rangle}{\langle\Psi|\Psi\rangle}(\bm{r},t^k)\left| i\dfrac{\partial\log\Psi}{\partial t}(\bm{r},t^k) - E_L(\bm{r},t^k) \right|^2 \dif{\bm{r}}\\
    & \quad + \dfrac{1}{N_t}\sum_{k=1}^{N_t}\int_{\Omega}\dfrac{\Psi^*\Psi}{\langle\Psi|\Psi\rangle}(\bm{r},t^k)(\nabla_\eta\left| i\dfrac{\partial\log\Psi}{\partial t}(\bm{r},t^k) - E_L(\bm{r},t^k) \right|^2)\dif{\bm{r}}\\
    & = \mathbb{E}_{p(\bm{r},t)}\left[ (\nabla_\eta\log\Psi^* + \nabla_\eta\log\Psi) \cdot \left| i\dfrac{\partial\log\Psi}{\partial t} - E_L\right|^2 \right] \\
    & \quad - \mathbb{E}_{p(t)} \left[\mathbb{E}_{p(\bm{r}\vert t)}\left[\nabla_\eta\log\Psi^* + \nabla_\eta\log\Psi\right](t^k) \cdot \mathbb{E}_{p(\bm{r}\vert t)}\left[\left| i\dfrac{\partial\log\Psi}{\partial t} - E_L\right|^2\right](t^k)\right]\\
    & \quad + \mathbb{E}_{p(\bm{r},t)}\left[\nabla_\eta\left| i\dfrac{\partial\log\Psi}{\partial t} - E_L\right|^2\right]\\
    \end{aligned}
\end{equation}
Here, $\mathbb{E}_{p(\bm{r},t)}$ denotes the expectation with respect to the probability distribution $p(\bm{r},t) = \dfrac{\Psi^*\Psi}{\langle\Psi|\Psi\rangle}$, and we decompose the joint distribution using the identity $\mathbb{E}_{p(\bm{r},t)} = \mathbb{E}_{p(t)} \mathbb E_{p(\bm{r}\vert t)}$, assume $\mathbb{E}_{p(t)}$ is uniform, and $\mathbb{E}_{p(\bm{r}\vert t)}$ indicates the expectation over the spatial domain at a specific time $t$.
The time-dependent local energy $E_L(\bm{r},t)$ can be computed analogously to the time-independent case, as the Hamiltonian operator does not contain time derivatives:
\begin{equation}
    \begin{aligned}
    E_L(\bm{r}, t) & =\Psi^{-1}(\bm{r},t) \hat{H} \Psi(\bm{r},t) \\
    & =-\frac{1}{2} \sum_j\left[\left.\frac{\partial^2 \log\Psi}{\partial r_j^2}\right|_{\bm{r},t}+\left(\left.\frac{\partial \log\Psi}{\partial r_j}\right|_{\bm{r},t}\right)^2\right]+V(\bm{r},t)
    \end{aligned}
\end{equation}
Here the index $j$ runs over all dimensions of the electron position vector; for example, in a 3D system with $N$ electrons, it ranges from 1 to $3N$.

\section{1D trapped interacting fermions}
\label{sec:1d-trapped-interacting-fermions}
First, we prepare a ground state in the potential $V(\bm{r})=\frac{\omega_0^2}{2} \sum_i r_i^2+g \sum_{i<j}\left(r_i-r_j\right)^2$. 
Under this time-independent potential, using the orthogonal coordinate transformation method from \cite{zaluska2000soluble}, we can derive the ground state energy as $E_0 = \frac{1+ (N^2-1)\omega}{2}$, where $N$ is the number of particles and $\omega = \sqrt{1+Ng^2}$. 
The ground state wavefunction (with the space-time separation phase) is given by
\begin{equation}
    \Phi(\bm{r}, t)=e^{-iE_0t}\mathcal{V}(\bm{r}) e^{-\frac{\omega}{2} \sum_i r_i^2-\frac{1-\omega}{2N}\left(\sum_i r_i\right)^2}
\end{equation}

At $t=0$, we perform a quench on the trap. 
The global Hamiltonian can be written as
\begin{equation}
    \begin{aligned}
    & V(\bm{r}, t<0)=\frac{\omega_0^2}{2} \sum_i r_i^2+g \sum_{i<j}\left(r_i-r_j\right)^2 \\
    & V(\bm{r}, t \geq 0)=\frac{\omega_f^2}{2} \sum_i r_i^2+g(t) \sum_{i<j}\left(r_i-r_j\right)^2
    \end{aligned}
\end{equation}
where $\omega_0=1$ and $\omega_f=2$ are the initial and post-quench trap frequencies, $g=1$ is the initial interaction strength, and $g(t)$ is the post-quench interaction strength, which will be determined during the construction of the analytical solution. 
Using the scaling method \cite{gritsev2010scaling}, we can solve the TDSE under this time-dependent Hamiltonian as
\begin{equation}
    \Psi\left(\bm{r}, t\right)=\frac{1}{R(t)} \mathrm{e}^{\mathrm{i} F(t) \sum_{i=1}^N r_i^2 }\Phi\left(\bm{y}, \tau\right)
\end{equation}
Here, $\bm{y}= \dfrac{\bm{r}}{L(t)}$ represents the scaled spatial coordinates, and $\tau(t) = \int_0^t \dfrac{ds}{L^2(s)}$ is the scaled time. 
By solving the Ermakov equation $\ddot{L}(t) +\omega^2(t) L(t)=\frac{\omega_0^2}{L^3(t)}$, we obtain the scaling function as
\begin{equation}
    L(t)=\sqrt{A \cos \left(2 \omega_f t\right)+C}, \quad A=\frac{\omega_f^2-\omega_0^2}{2 \omega_f^2}, \quad C=\frac{\omega_f^2+\omega_0^2}{2 \omega_f^2}
\end{equation}

From this, we can derive the scaled time as
\begin{equation}
    \tau(t) = \frac{1}{\omega_0} \arctan\left( \frac{\omega_0}{\omega_f} \tan(\omega_f t) \right) + \frac{\pi}{\omega_0} \cdot \left\lfloor \frac{\omega_f t}{\pi} + \frac{1}{2} \right\rfloor
\end{equation}

The post-quench interaction strength function, the phase-related function $F$, and the normalization constant function $R$ are all derived from the scaling function $L$ as follows
\begin{equation}
    \begin{aligned}
    & g(t) = \frac{g}{L^4(t)} \\
    & F(t) = \frac{\dot{L}(t)}{2L(t)} \\
    & R(t) = L^{\frac{ND}{2}}(t)
    \end{aligned}
\end{equation}
The monopole breathing mode behavior of the model is also described by the scaling function. 
The time evolution of the monopole is given by
\begin{equation}
    \begin{aligned}
        M(t) &= \sum_{i=1}^N \langle r_i^2 \rangle_{t} = \sum_{i=1}^N \int \Psi^*(\bm{r},t)\Psi(\bm{r},t) r_i^2 \dif{\bm{r}} \\
        &= \sum_{i=1}^N \dfrac{1}{R^2(t)} \int \psi^2(\frac{\bm{r}}{L(t)})r_i^2 \dif{\bm{r}} \\
        &= \sum_{i=1}^N \dfrac{L^{N+2}(t)}{R^2(t)} \int \psi^2(\frac{\bm{r}}{L(t)})(\dfrac{r_i}{L(t)})^2 \dif{\bm{r}} \\
        &= L^2(t) \sum_{i=1}^N \langle x_i^2 \rangle_{t=0} = L^2(t)M(0)
    \end{aligned}
\end{equation}
where $M(0)$ is the initial monopole moment, which we calculate numerically in practice.

\section{Computational Resources and Cost}
\label{sec:computational-resources-and-cost}

We report the wall-clock time for the simulation of H$_2$ dissociation in a laser field described in Section 3.5.

The FASTNet calculations were performed on a high-performance computing node equipped with 8 $\times$ NVIDIA A800 GPUs. 
The simulation spanned 44 time windows, and the total cumulative wall-clock time for the full pulse propagation was approximately 250 hours. 
For comparison, standard Time-Dependent Full Configuration Interaction (TD-FCI) calculations were performed on a consumer workstation (Apple M2 Max CPU) using the PySCF package. 
The reference calculation took 0.19 seconds using the STO-3g basis set and 14.12 seconds using the cc-pVDZ basis set.

We acknowledge that for the small system studied here ($N=2$), the machine learning approach is orders of magnitude more computationally expensive than optimized basis-set methods. 
This high cost stems primarily from the fixed overhead of variational Monte Carlo sampling and the extensive optimization required for deep neural networks at every time step. 
However, this cost should be contextualized within the scaling limits of the respective methods. 
Standard FCI scales exponentially with electron number, whereas FASTNet scales polynomially. 
While we currently lack a speed advantage for few-electron systems, the polynomial scaling offers a potential pathway to study larger, strongly correlated systems. 
Reducing the computational burden remains a critical challenge and a priority for our future work.

\end{document}